\documentclass{aa}  
\usepackage{float}
\usepackage{multirow}

\usepackage{lineno}

\renewcommand{\linenumbers}{}

\usepackage{graphicx}
\usepackage{txfonts}

\usepackage{subcaption}
\usepackage{lscape}             
\usepackage{placeins}

\usepackage{orcidlink}

\hypersetup{colorlinks=true, citecolor=blue, linkcolor=blue, urlcolor=blue}

\RequirePackage{xcolor}

\RequirePackage{listings}
\lstdefinelanguage{DIFcode}{moredelim=[il][\color{white}\tiny]{\%DIF\ <\ },moredelim=[il][\sffamily\bfseries]{\%DIF\ >\ }
} 
\lstdefinestyle{DIFverbatimstyle}{
        language=DIFcode, 
        basicstyle=\ttfamily, 
        columns=fullflexible, 
        keepspaces=true 
} 
\lstnewenvironment{DIFverbatim}{\lstset{style=DIFverbatimstyle}}{}
\lstnewenvironment{DIFverbatim*}{\lstset{style=DIFverbatimstyle,showspaces=true}}{} 
\begin{document}

        \title{The $m_\mathrm{max}-M_\mathrm{ecl}$ relation in the LEGUS clusters}

        \author{Marie Zinnkann\inst{1} \and Tereza Jerabkova\inst{2} 
                \and Zhiqiang Yan\inst{3,4} \and Pavel Kroupa \inst{1,5}  \and Yannik Ostermann\inst{1} \and Eda Gjergo\inst{3,4}
                \and Akram Hasani Zonoozi\inst{1,6} \and Hosein Haghi\inst{1,6,7} \and Jan Pflamm-Altenburg\inst{1}}
        
        \institute{Helmholtz-Institut für Strahlen- und Kernphysik, Universität Bonn, Nussallee 14-16, 53115 Bonn, Germany, \\ \email{mzinnkann@astro.uni-bonn.de}
                \and Department of Theoretical Physics and Astrophysics, Faculty of Science, Masaryk University, Kotl\'{a}\v{r}sk\'{a} 2, Brno 611 37, Czech Republic,  
                \email{tereza.jerabkova@sci.muni.cz}
                \and School of Astronomy and Space Science, Nanjing University, Nanjing 210093, PR China 
                \and Key Laboratory of Modern Astronomy and Astrophysics (Nanjing University), Ministry of Education, Nanjing 210093, PR China 
                \and Astronomical Institute, Faculty of Mathematics and Physics, Charles University, V Holešovickách 2, 180 00 Praha 8, Czech Republic
                \and Department of Physics, Institute for Advanced Studies in Basic Sciences (IASBS), PO Box 11365-9161, Zanjan, Iran \and School of Astronomy, Institute for Research in Fundamental Sciences (IPM), PO Box 19395 - 5531, Tehran, Iran
        }
        
        \date{Received \today}

        \abstract 
        {The relation between the maximum stellar mass in a very young cluster ($m_\mathrm{max}$) and the total stellar mass of the cluster ($M_\mathrm{ecl}$), known as the $m_\mathrm{max}-M_\mathrm{ecl}$ relation, remains debated in the literature. To test the validity of this relation, we modelled young star clusters with masses between $10^{2.5}$ and $10^{5.0} M_\odot$ and ages of $1 - 4\,\mathrm{Myr}$ using the \texttt{galIMF} code, in which stellar masses are optimally sampled from a varying initial stellar mass function. We compared the results with literature observations of extragalactic young star clusters.
                We incorporated stellar evolution via \texttt{PARSEC} and \texttt{COLIBRI} tracks and computed  H$\alpha$ luminosities using the \texttt{P\'egase} code.
                To account for dynamical ejections, we stochastically removed stars based on their spectral type, following previous $N$-body simulations.
                Additional sources of scatter, including uncertainties in age determination and contamination by field stars, were considered. 
                Our results indicate that, under the assumptions explored here, optimal sampling is consistent with the extragalactic star cluster observations considered, whereas purely random sampling produces distributions that are not in agreement. These findings support a highly self-regulated interpretation of cluster formation in which stellar masses align optimally with the initial mass function rather than being drawn independently at random.}        
        
        \maketitle
\section{Introduction}\label{sec:intro}
Whether stars in stellar systems form through stochastic sampling or through ordered, physically regulated processes has direct consequences for how stellar populations, cluster demographics, and feedback are interpreted. Most analytic models and population synthesis approaches implicitly assume that stellar masses are drawn randomly from a universal initial mass function (IMF; \citealt{Kroupa_VariationOnIMF, Chabrier_IMF2003}). However, observational and theoretical work challenges this view. \citet{Weidner_mmax_Mecl_2013} and \citet{Yan_mmax_greenplot} show that the observed relation between the stellar mass of the embedded star cluster, $M_\mathrm{ecl}$, and the maximum stellar mass, $m_\mathrm{max}$,  is inconsistent with purely random sampling, indicating physical regulation during cluster formation. The relation between $m_\mathrm{max}$ and the mass of the molecular cloud clump was noticed by \citet{Larson_mmax_Mcloud_relation} and formulated as a power-law relation in \citet{Larson_mmax_Mecl_relation}.
 The existence of a well-defined relation between $m_\mathrm{max}$ and $M_\mathrm{ecl}$  for clusters born in molecular clumps was quantified for the first time by \citet{Weidner_mmax_Mecl_2006}.
They showed with multiple Monte Carlo experiments that observational data of young clusters are in better agreement with an ordered formation than a random one.
  This has been interpreted as evidence of deterministic IMF sampling, of which optimal sampling is a specific case \citep{Kroupa_IMF_history, Schulz_optimalsampling}.

Optimal sampling is a specific deterministic scheme in which stellar masses are assigned so that the IMF is matched exactly in all mass intervals. Deterministic sampling refers to any sampling scheme in which a given IMF and total stellar content uniquely and reproducibly determine the resulting set of stellar masses, implying a highly regulated form of star formation with no intrinsic sampling noise.
Random sampling treats the IMF as a probability distribution from which stellar masses are drawn independently, implying that star formation is intrinsically stochastic and that finite stellar populations exhibit unavoidable statistical fluctuations about the IMF \citep{Kroupa_densepopulations_2008}. \citet{Weidner_mmax_Mecl_2006} ruled out random sampling with $99.9\%$ confidence, which was confirmed in \citet{Weidner_mmax_Mecl_2013}. 
Using observational data of the young star-forming regions Taurus, Lupus3, ChaI, and IC348 \citep{Kirk_mmax_Mecl}, young stellar clusters in the Large Magellanic Cloud \citep{Stephens_mmax_Mecl}, the star-forming regions Taurus-Auriga and L1641 in Orion \citep{Weidner_mmax_Mecl_2013}, and new observational data from the VVV survey \citep{Ramirez_mmax_Mecl}, \citet{Yan_GalIMFCode2017} show that the $m_\mathrm{max}$--$M_\mathrm{ecl}$ relation is consistent with deterministic sampling approaches. Random sampling can also reproduce the overall trend, but typically with a larger intrinsic scatter than observed \citep{Yan_mmax_greenplot}. Furthermore, deterministic and optimal sampling are consistent with a highly self-regulated picture of star formation \citep{Yan_GalIMFCode2017,Vazquez_starformation}.

The sampling method also affects our understanding of galaxies whose stellar content is the sum of the stars in the galaxy's embedded star clusters, in which the majority of stars form \citep{Kroupa_FieldstarIMF}. This concept is included in the theory of the integrated galaxy-wide IMF, which comprises the distribution of stellar remnants, the number of supernovae, and the chemical enrichment history of a galaxy \citep{Weidner_IGIMF, Jerabkova_metallicity_influence, Haslbauer_varyingIMF, Zonoozi_varyingIMF}. This theory also has several cosmological implications, which are listed in \citet{Kroupa_2026}, and it solves multiple problems, such as the short star formation timescales of elliptical galaxies, named the downsizing problem \citep{Yan_depending_IMF2021}, with significant implications for our understanding of the cosmic microwave background \citep{Gjergo_CMB}. 

Despite evidence of the $m_\mathrm{max}$--$M_\mathrm{ecl}$ relation across multiple associations and spatial scales, several studies have challenged it (e.g. \citealt{Andrews_NGC4214, Jung_LEGUS}).  \citet{Weidner_mmax_Mecl_2013} argued that the reported contradictions can arise from the interpretation of the observational data given observational biases. By revisiting the analysis of \citet{Andrews_NGC4214}, \citet{Weidner_NGC4214} defended the {$m_\mathrm{max}$--$M_\mathrm{ecl}$} relation.
The present work   {revisits the interpretation presented by \citet[hereafter J23]{Jung_LEGUS}. In their analysis, the authors   {report no clear evidence for the $m_\mathrm{max}$--$M_\mathrm{ecl}$ } relation in their observed megaparsec-distant galaxies of the Legacy ExtraGalactic UV Survey (LEGUS). Young,  massive stars emit high-energy photons that ionise the surrounding gas. Hydrogen recombination produces a prominent H$\alpha$ emission line, which is therefore a useful tracer of strongly ionising stellar populations \citep{PflammAltenburg_LHalpha}.
Assuming a canonical IMF and a maximum stellar mass of $120\,M_\odot$, the approach by J23 was to analyse the dependence of the upper end of the IMF on the cluster mass by normalising the measured $\mathrm{H}\alpha$ luminosity, $L_{\mathrm{H}\alpha}$, by $M_\mathrm{ecl}$ and plotting it against $M_\mathrm{ecl}$\footnote{By using clusters with observationally derived ages of about $4\,\mathrm{Myr}$, we can reasonably assume the observationally deduced mass in stars in the observed clusters, $M_\mathrm{cl}$, to be similar to $M_\mathrm{ecl}$.}. The result is displayed in   {their Fig.~4.
In that diagnostic plot, the clusters appear broadly consistent with a randomly populated distribution, from which J23 conclude that the {$m_\mathrm{max}$--$M_\mathrm{ecl}$ relation is not evident in their observed galaxies.

The goal of this work is to test which cluster-sampling process is most consistent with observations. We optimally and randomly sampled star clusters and determined their H$\alpha$ luminosities for both sampling cases. We then applied evolutionary processes (stellar mass evolution and dynamical ejections) and considered observational uncertainties (age uncertainty and contamination by field stars). Finally, we compared the resulting cluster distributions to the observations.

The concepts that are relevant for this work are described in Sect.~\ref{sec:Methods}. The results are shown in Sect.~\ref{sec:Results} and are discussed in Sect.~\ref{sec:Discussion}. This work is summarised in Sect.~\ref{sec:Conclusion}.

\section{Methods}\label{sec:Methods}
In this section we describe the framework used to model the stellar content and the resulting H$\alpha$ luminosity of young star clusters. The primary goal is to predict the ionising output of a cluster of given mass and age by explicitly accounting for the stochastic and systematic effects associated with sampling the stellar IMF and subsequent evolutionary processes. The model combines a variable IMF prescription, different IMF sampling schemes, stellar evolution–based mass and luminosity estimates, and a treatment of observational and physical uncertainties. Together, these ingredients allow us to quantify how cluster mass, age, and IMF sampling affect the inferred H$\alpha$ emission and the observable stellar population.
 {Twenty embedded cluster masses, $M_{\rm ecl}$, were chosen uniformly distributed over log-mass in the range $10^{2.5}\,M_\odot$ to $10^{4.5}\,M_\odot$. We note here that we merely needed to cover the approximate range of clusters covered observationally by J23 but with a larger mass range to account for clusters that move out of the observational range due to evolutionary processes and the biases studied here. The ensemble of embedded cluster masses thus comprises the following values in $M_\odot$: 316, 428, 579, 784, 1062, 1438, 1947, 2636, 3569, 4832, 6543, 8858, 11993, 16237, 21983, 29763, 40296, 54555, 73861, and 100000. For the final comparison to J23, we restricted the model to the mass range covered by the observed sample and down-sampled the highest-mass model clusters to reflect the fact that J23 included only a few clusters at the high-mass end.
}

\subsection{The IMF}\label{sec:meth_IMF}
The initial stellar mass function was introduced as a one-part   {power law, $\mathrm{d} N = \xi(m)\,\mathrm{d}m = k m^{-\alpha}\,\mathrm{d}m$}, where $\mathrm{d} N$ is the number of stars in the mass interval between $m$ and $m+\mathrm{d}m$, $k$ is a normalisation constant, and $\alpha$ is the slope of the mass function. \citet{Salpeter_IMF} derived the IMF in the mass range $0.4 \leq m/M_\odot \leq 10$ with a slope of $\alpha = 2.35$ from the luminosity function of main-sequence stars in the solar neighbourhood.

Based on an extensive star-count analysis of the Galactic field by \citet{KTG_93} and observations of young star clusters and OB associations in the Milky Way and the Magellanic Clouds, \citet{Kroupa_IMF_variations} concluded that the IMF is a multi-part power law; the data support a three-part power law (the canonical IMF):
\begin{align}
        \xi(m) = k
        \begin{cases}
                m^{-\alpha_1}, & m_\mathrm{min} < m \leq m_1, \\
                k_2 \cdot m^{-\alpha_2}, & m_1 < m \leq m_2, \\
                k_3 \cdot m^{-\alpha_3}, & m_2 < m \leq m_\mathrm{max} \, ,
        \end{cases}
\end{align}
where $k_i$ has to be determined such that the IMF fulfils continuity and $k$ is needed for normalisation.
Furthermore, $m_\mathrm{min}=0.08\,M_\odot$ and $m_\mathrm{max}$ are the lowest and maximum stellar mass, respectively. 
Studying the rich star clusters R136 and Arches, masses up to $150\,M_\odot$ could be observed \citep{Weidner_masslimit_2004, Figer_mmax}, which will be adopted in this work as the maximum physical stellar mass. More massive stars are likely mergers \citep{Banerjee_mergers} and brown dwarfs appear as objects associated with the formation of stars (\citeauthor{Kroupa_2026} \citeyear{Kroupa_2026} for further information). In the canonical IMF, $\alpha_1=1.3$ and $\alpha_2=\alpha_3=2.3$.

The systematic change of the IMF is observable in extreme star-forming environments and metallicities different compared to the Milky Way's as shown in \citet{Dabringhausen_dSph, Dabringhausen_topheavy_dSph, Dabringhausen_XRayBinaries}.
\citet{Marks_varying_IMF} found a `fundamental plane' that describes the change of the IMF with density and metallicity. The authors also find that the density of the molecular cloud clump has a bigger impact on the upper end of the IMF than the metallicity. 

In \citet{Yan_GalIMFCode2017},  Eqs.~(\ref{eq:varying_IMF_alpha1}) and (\ref{eq:varying_IMF_alpha2}) were empirically determined and $\Delta \alpha = 63$ is suggested for $Z_\odot = 0.0142$ \citep{Yan_depending_IMF2021},  
\begin{align}
        \alpha_1 &= 0.3 + \Delta \alpha \cdot (\mathrm{Z} - \mathrm{Z}_\odot)  \,\,\,\,\,\,\,\,\,\,\,\,\,\,\,\,\,\,\,\,\,\,\,\, 0.08 \leq m/M_\odot < 0.5 \, , \label{eq:varying_IMF_alpha1}\\ 
        \alpha_2 &= 1.3 + \Delta \alpha \cdot (\mathrm{Z} - \mathrm{Z}_\odot)  \,\,\,\,\,\,\,\,\,\,\,\,\,\,\,\,\,\,\,\,\,\,\,\, 0.5 \leq m/M_\odot < 1.0 \, . \label{eq:varying_IMF_alpha2}
\end{align}
The dependence of the power-law index for high-mass stars was determined empirically by \citet{Marks_varying_IMF}. The formulation by \citet{Yan_GalIMFCode2017} that is used in this work is
\begin{align} \label{eq:varying_IMF_alpha3} 
        \alpha_3 &= 
        \begin{cases}
                2.3,  \,\,\,\,\,\,\,\,\,\,\,\,\,\,\,\,\,\,\,\,\,\,\,\,\,\,\,\,\,\,\,\,\,\,\,x < -0.87 \\
                - 0.41x + 1.94, \,\,\,\,\,\,\,\,x > -0.87
        \end{cases} \,\,\,\,\,\, \,\,1 \leq \frac{m}{M_\odot} < m_\mathrm{max} \,, 
\end{align}

\noindent with $x = -0.14 [\mathrm{Z}] + 0.99\log_{10} (\rho_{\text{cl}} / (10^6\text{M}_\odot\text{pc}^{-3}))$, where $\rho_{\text{cl}}$ is the clump density (final stars plus residual gas assuming a star formation efficiency of 1/3) and $[Z] = \log_{10} (Z/Z_\odot) \approx [Z/X]$ with the initial metal-mass fraction, $Z$,  and hydrogen-mass fraction, $X$, in stars.

\subsection{Sampling of star clusters}\label{sec:sampling_methods}
To obtain a finite stellar population, the IMF must be discretised. Here we considered two widely used discretisation schemes: random sampling (Sect.~\ref{sec:random_sampling}) and optimal sampling (Sect.~\ref{Optimal_sampling}). In optimally sampled embedded clusters, stellar masses are assigned such that the discretised population matches the IMF exactly in each mass interval; by construction this minimises sampling noise and avoids gaps \citep{Kroupa_IMF_history, Yan_GalIMFCode2017}. In contrast,  randomly sampled clusters treat the IMF as a probability distribution and therefore show finite-population fluctuations (including gaps), and can occasionally contain more massive stars at fixed $M_\mathrm{ecl}$ compared to the optimal-sampling case \citep[see Fig.~1 of][]{Kroupa_IMF_history}.

\subsubsection{Random sampling}\label{sec:random_sampling} 
For random sampling, the IMF is treated as a probability density function, $p(m)\propto \xi (m)$. Using a generating function, stellar masses can be drawn randomly from the IMF as described in \citet{Kroupa_densepopulations_2008}.
First, the normalisation constant needs to be determined using the following condition:
\begin{align*}
        1 = \int_{m_\mathrm{min}}^{m_\mathrm{max}} p (m) \, \mathrm{d}m \, .
\end{align*}
From the continuity condition $m_1^{-\alpha_1} = k_2m_2^{-\alpha_2}$ and $k_2m_2^{-\alpha_2} = k_3m_3^{-\alpha_3}$, $k_2$ and $k_3$ can be determined.

After defining the generating functions with the IMF slopes in Eqs.~(\ref{eq:varying_IMF_alpha1}), (\ref{eq:varying_IMF_alpha2}), and (\ref{eq:varying_IMF_alpha3}) and a random number between 0 and 1, the corresponding stellar mass was obtained as described in \citet{Kroupa_densepopulations_2008}.
Based on this concept, we developed a \texttt{Python} code\footnote{\url{https://github.com/MarieZkn/Random_sampling}} that draws stellar masses until the target $M_\mathrm{ecl}$ is reached, and then removes the final draw to avoid exceeding the target mass. Consequently, the realised stellar mass is slightly below the initial target $M_\mathrm{ecl}$; throughout the analysis we therefore used the realised cluster mass, computed by summing the drawn stellar masses.

\subsubsection{Optimal sampling}\label{Optimal_sampling}
In optimal sampling, the IMF is assumed to be an optimal distribution function where the number of stars in any stellar mass range matches exactly the integration of the given IMF in that mass range  \citep{Kroupa_IMF_history}.
Codes for optimal sampling are provided by \citet{Yan_GalIMFCode2017}\footnote{\url{https://github.com/Azeret/galIMF}}   and \citet{Gjergo_pyIGIMF}\footnote{\url{https://github.com/egjergo/pyIGIMF}}, based on the mathematical formulation of optimal sampling by \citet{Schulz_optimalsampling}.

To optimally sample the stellar IMF in an embedded cluster, the normalisation constant and $m_\mathrm{max}$ need to be determined with the following conditions: the mass conservation law $M_\mathrm{ecl} = \int_{m_\mathrm{min}}^{m_\mathrm{max}} m\,\xi (m)\, \mathrm{d}m$ and the optimal-sampling normalisation condition $1 = \int_{m_\mathrm{max}}^{150\,M_\odot}\xi(m)\,\mathrm{d}m$.
The different integral boundaries, $m_i$, are determined by summing up all stars as described in Eq.~(\ref{eq:N_tot_optimal}), where each integral must be one and $N_\mathrm{tot}$ is the calculated total number of stars, 
\begin{align}
        N_\mathrm{tot} &= \int_{m_\mathrm{min}}^{m_\mathrm{max}} \xi(m)\,\mathrm{d}m \notag \\
        &= \int_{m_2}^{m_\mathrm{max}} \xi(m)\,\mathrm{d}m + ... + \int_{m_{N_\mathrm{tot}}}^{m_{N_\mathrm{tot}-1}} \xi(m)\,\mathrm{d}m + \int_{m_\mathrm{min}}^{m_{N_\mathrm{tot}}} \xi(m)\,\mathrm{d}m \,. \label{eq:N_tot_optimal}
\end{align}
The calculation starts with the most massive star. So with $\int_{m_2}^{m_\mathrm{max}} \xi(m)\,\mathrm{d}m = 1$ and the before deduced $m_\mathrm{max}$, $m_2$ can be calculated. Then, the remaining $m_i$ are determined iteratively. With $m_i = \int_{m_{i+1}}^{m_i} m \, \xi(m)\,\mathrm{d}m$, the corresponding mass of the star is calculated.

\subsection{Stellar evolution}\label{sec:stellar_evolution}
We assumed that all stars in an embedded cluster have the same initial age. We accounted for stellar mass evolution with time using evolutionary tracks\footnote{downloaded from \url{http://stev.oapd.inaf.it/cmd}}. They were selected to start from the pre-main sequence and continue until either the first thermal pulsing or C-ignition from \texttt{PARSEC} \citep{Bressan_PARSEC} version 1.2S, which is available for $0.0001\leq Z \leq 0.02$ in the mass range $0.1\leq m/M_\odot < 350$, for $0.03\leq Z\leq 0.04$ in the mass range $0.1\leq m/M_\odot < 150,$ and for $Z=0.06$ in the mass range $0.1\leq m/M_\odot<20$ (cf. \citeauthor{Tang_PARSEC} \citeyear{Tang_PARSEC} for $0.001\leq Z \leq 0.004$ and \citeauthor{Chen_PARSEC_2015} \citeyear{Chen_PARSEC_2015} for other $Z$), with revised and calibrated surface boundary conditions in low-mass dwarfs provided in \citet{Chen_PARSEC_2014}.
Furthermore, the thermal pulsing-asymptotic giant branch evolution from the first thermal pulse to the total loss of the envelope is added with \texttt{COLIBRI} S\_37 \citep{Pastorelli_Giantbranchphase_LMC} tracks for $0.008 \leq Z \leq 0.02$, \texttt{COLIBRI} S\_35 tracks \citep{Pastorelli_Giantbranchphase_SMC} for $0.0005 \leq Z \leq 0.006$ and \texttt{COLIBRI} PR16 \citep{Marigo_PARSE_CCOLIBRI} for $Z\leq0.0002$ and $Z\geq 0.03$.

The tracks were downloaded for ages between 1 and 10 Myr and $Z=\{0.004,\, 0.008,\, Z_\odot\approx 0.014\},$ where 0.014 is the solar metallicity according to \citet{Asplund_Z_sun}. With interpolation, the mass of a star at a given age, $m$, is connected to the initial mass, $m_\mathrm{ini}$.

\subsection{H$\alpha$ luminosity}\label{subsec:Pegase}
The H$\alpha$ luminosity was computed using the galaxy evolution code \texttt{P\'egase} \citep{Fioc_Pegase2, Fioc_Pegase3}, which models the spectra and chemical evolution of galaxies across a wide wavelength range\footnote{For more details and access to the code, see \url{https://www2.iap.fr/users/fioc/PEGASE.html}}.
We computed synthetic spectra by setting up simple stellar populations (SSPs), where all stars have the same mass and age. The technical details are described in Appendix~\ref{sec:App_Halpha_1}.
The obtained H$\alpha$ luminosity of the SSP was normalised to the one of a single star. 
To do so, the number of objects in the SSPs, $N_\mathrm{tot}$, needed to be obtained with the given IMF and the given stellar mass, which was normalised to $1\,M_\odot$ by \texttt{P\'egase}.   

\subsection{Dynamical ejections} 
\label{subsec:Dynamical_ejections}
For clusters younger than $4\,\mathrm{Myr}$, gas expulsion has not significantly changed $M_\mathrm{ecl}$ \citep{Weidner_relation_mmax_parent}, but the dynamical ejection of stars may have an impact. Depending on the initial radius, clusters can eject their most massive stars within the first {$3\,\mathrm{Myr}$ \citep{Oh_influence_of_dynamical_ejections}. To assess the influence of dynamical ejections on the {$m_\mathrm{max}$--$M_\mathrm{ecl}$ relation, we adopted the results of \citet{Oh_Dependency_DynamicalEjections} and \citet{Oh_influence_of_dynamical_ejections, Oh_Kroupa_Dynamicalejections_initial_conditions}.

 \citet{Oh_influence_of_dynamical_ejections} performed \texttt{NBODY6} \citep{Aarseth_NBODY6} simulations of young star clusters with masses in the range $M_\mathrm{ecl} = 10-10^{3.5}M_\odot$, half-mass radii of $r_\mathrm{h} = \{0.3, 0.8\}\,\mathrm{pc,}$ and binary fractions of 0 (all single stars) and 1 (all binaries). The binaries were randomly paired or ordered paired; the latter means that stars heavier than $5\,M_\odot$ were paired to binaries with mass ratios close to unity. Lower-mass stars were randomly paired (for a discussion, see \citeauthor{Kroupa_2026} \citeyear{Kroupa_2026} and \citeauthor{Kroupa_2025} \citeyear{Kroupa_2025}).
In \citet{Oh_influence_of_dynamical_ejections}, and the positions and velocities of stars in the clusters were generated according to the Plummer model, which is explained in \citet{Aarseth_Plummer_model}. 
For stellar evolution, the stellar evolution library \citep{Hurley_NBODY6_stellar_evolution} in the \texttt{NBODY6} code was used. \citet{Oh_Kroupa_Dynamicalejections_initial_conditions} analysed the simulations performed in \citet{Oh_influence_of_dynamical_ejections} and \citet{Oh_Dependency_DynamicalEjections} for $M_\mathrm{ecl} = 10^{4}\,M_\odot$ and $10^{4.5}\,M_\odot$.

Referring to \citet{Goodwin_stellarmultiplicty}, most stars form in systems of two or three stars. 
Furthermore, \citet{Sana_binary_fraction} studied O-type stars in NGC 6231, finding that random pairing is unlikely for O-type stars and that O+OB binaries are favoured. This is confirmed in \citet{Sana_binary_fraction2} and other publications. 
Furthermore, star clusters are commonly initially mass-segregated   {\citep{Bonnell_masssegregation, Gouliermis_Masssegregation_MC, Chen_masssegregation, Pavlik_masssegregation}, which is accounted for in the simulations by \citet{Oh_influence_of_dynamical_ejections} and \citet{Oh_Dependency_DynamicalEjections}}.

\citet{Marks_rh_Mecl} performed an inverse dynamical population synthesis. By studying binary populations together with constraints on the birth densities of globular clusters obtained in \citet{Marks_GCinitialconditions}, a correlation between $M_\mathrm{ecl}$ and the half-mass radius $r_\mathrm{h}$ of the form $\frac{r_\mathrm{h}}{\mathrm{pc}} = 0.10^{+0.07}_{-0.04} \cdot \Big(\frac{M_\mathrm{ecl}}{M_\odot}\Big)^{0.13\pm0.04}$ was found.
With this, $r_\mathrm{h}$ of the clusters considered in this analysis is estimated to be $0.3\,\mathrm{pc}$ within $1\sigma$ confidence.
  We therefore adopted model \texttt{MS3OP} of \citet{Oh_Dependency_DynamicalEjections}, which assumes initial mass segregation, a half-mass radius of $r_\mathrm{h}\approx 0.3\,\mathrm{pc}$, and ordered pairing for massive binaries.

The ejection fraction depends on the spectral type and, hence, stellar mass \citep[Fig.~1]{Oh_Kroupa_Dynamicalejections_initial_conditions}. The division of the spectral types is displayed in Table~\ref{tab:Ejection_Properties}. 
The mean ejection fraction of O-star systems, i.e. of multiple stars, at an age of $3\,\mathrm{Myr}$, $\langle f_\mathrm{ej,O-sys} \rangle = \frac{N_\mathrm{ej,O-sys}}{N_\mathrm{O-sys}}$, with the number of ejected O-star systems, $N_\mathrm{ej,O-sys}$, and the number of O-star systems in the cluster, $N_\mathrm{O-sys}$, is listed in Table~\ref{tab:sigma_values} according to \citet[their Table~A1]{Oh_Dependency_DynamicalEjections}. We note that the notation used in this work differs slightly from that in the original reference.
A star or star system is classified as ejected  if its distance from the cluster centre is greater than $3\times r_\mathrm{h}$ of the cluster at   {$3\,\mathrm{Myr}$}, and its velocity is larger than the escape velocity at that radius.
With a Gaussian function of the form $f_\mathrm{ej,O-sys} \approx 0.23 \cdot \exp\Big(-\frac{1}{2} (\frac{\log(M_\mathrm{ecl}/M_\odot)-3.5}{-0.70})^2\Big)$, the ejection fractions of O-star systems for different cluster masses were inferred. 
Furthermore, the uncertainties were taken into consideration by adding the values of the standard deviation, which are listed in Table~\ref{tab:sigma_values} and were determined from the values of all data points, and calculating the standard deviation. In the case of multiple points with the same value where the determination of the number of points is difficult, the number is estimated to yield the number of data points ($N_\mathrm{run}$ in Table~\ref{tab:sigma_values}) and the mean value to be in agreement with $\langle f_\mathrm{ej,O-sys}\rangle$ in Table~\ref{tab:sigma_values}. 
In order to determine $\sigma_{\langle f_\mathrm{ej,O-sys}\rangle}$ for a $M_\mathrm{ecl}$ that is not listed in Table~\ref{tab:sigma_values}, the value of the closest $M_\mathrm{ecl}$ for which a value for $\sigma_{\langle f_\mathrm{ej,O-sys}\rangle}$ is listed is assumed.
Thus, the ejection fraction can be increased within $\sigma$ confidence regions. In the case of a $2\sigma$ higher ejection fraction, $\sigma_{\langle f_\mathrm{ej,O-sys}\rangle}$ is assumed to be twice as high.

\citet[]{Oh_Dependency_DynamicalEjections} present the averaged binary fraction of ejected O-star systems in their Fig.~11. The values are listed in the third column of our Table~\ref{tab:sigma_values}. To obtain the binary fraction, we fitted a function of the form $\langle f_\mathrm{b,ej, O} \rangle \approx 32.07 \cdot (M_\mathrm{ecl}/M_\odot) ^{-0.65}$ to the data points of model \texttt{MS3OP}. They found that the mass-ratio distribution does not significantly influence the ejection of massive stars as long as massive stars are paired with other massive stars in binaries. We therefore considered stars from the same spectral type for the ejection of systems but did not apply ordered pairing to the collection of ejected stars.

Assuming a constant ejection rate during ages from $1-3\,\mathrm{Myr}$, the fraction of ejections at a specific age $t$ can be determined with   
\begin{align*}
        f_\mathrm{t} = \begin{cases}
                (t/\mathrm{Myr})/3 \,\,\,\,\,\,\,\, &\mathrm{if}\,\,\, t \leq 3\,\mathrm{Myr} \, ,\\
                1 \,\,\,\,\,\,\,\, &\mathrm{if}\,\,\, t > 3\,\mathrm{Myr} \,.
        \end{cases}
\end{align*}
The number of ejected O stars is thus $N_\mathrm{ej,O} = N_\mathrm{ej,O-sys} \cdot (1 + f_\mathrm{b,ej, O}) \cdot f_\mathrm{t}$.
  Note that $f_\mathrm{ej, O-sys} = \frac{N_\mathrm{ej,O-sys}}{N_\mathrm{O-sys}}$ (not $\frac{N_\mathrm{ej,O-sys}}{N_\mathrm{O}}$). If all stars are in binaries,  then $N_\mathrm{O-sys}$ can be approximated as $N_\mathrm{O-sys} \approx \frac{1}{2} N_\mathrm{O}$.
So, the final ejection fraction of O-type stars is calculated as 
\begin{align*}
        f_\mathrm{ej,O} = \frac{N_\mathrm{ej,O}}{N_\mathrm{O}} = \frac{1}{2} \cdot f_\mathrm{ej, O-sys} \cdot (1 + f_\mathrm{b, ej, O}) \cdot f_\mathrm{t} \,.
\end{align*}

\noindent With the fraction conversion $f_\mathrm{conv}$ displayed in Table~\ref{tab:Ejection_Properties}, the ejection fractions of the other spectral types are determined.   {The values of $f_\mathrm{conv}$ are inferred from Fig.~1 of \citet{Oh_Kroupa_Dynamicalejections_initial_conditions}.
}

\subsection{Age uncertainty}\label{sec:Meth_age_uncertainty}
In this work we assumed that stars within a cluster have the same age. Observationally inferred cluster ages, however, can carry uncertainties, which propagate into the inferred stellar evolutionary state and (in our framework) into both the expected dynamical-ejection fraction and the H$\alpha$ luminosity. J23 discuss age uncertainties at the level of $\approx 1$--$2\,\mathrm{Myr}$}.

  To account for this, we treated the age uncertainty  as a sensitivity test by re-evaluating the stellar masses implied by a given ionising output at a shifted age. Concretely, we kept the previously computed $L_{\mathrm{H}\alpha}$ for each star and inferred the corresponding initial mass ($m_\mathrm{ini}$) using the $L_{\mathrm{H}\alpha}$ -- $m_\mathrm{ini}$ relation at the shifted age. We then used the evolutionary tracks (Sect.~\ref{sec:stellar_evolution}) to obtain the stellar mass at that age. When the mapping is not single-valued at older ages (multiple $m_\mathrm{ini}$ yielding the same $L_{\mathrm{H}\alpha}$), we chose the solution closest to the original mass to minimise discontinuous jumps. If $L_{\mathrm{H}\alpha}$ exceeded the maximum value produced at the shifted age in our precomputed grid, we treated this as an indication that the star must be younger than the shifted age within our modelling framework and therefore used the youngest age bin considered for the $L_{\mathrm{H}\alpha}$--$m_\mathrm{ini}$ mapping.

\subsection{Contamination by field stars}\label{sec:Meth_fieldstars} 
The pixel scale of the camera used by J23 is $0''.0396\,\mathrm{pixel}^{-1}$. J23 chose an aperture radius of 5 pixels, corresponding to $8$--$10\,\mathrm{pc}$ for the distances ($4.3$--$5.1\,\mathrm{Mpc}$) of the selected galaxies \citep{Jung_LEGUS}. The targeted clusters have diameters smaller than $9\,\mathrm{pc}$, and so crowded environments and aperture-based selection can lead to residual local contamination and imperfect membership assignment.

We accounted for this using a controlled sensitivity test and adding additional stars to each model cluster. The added stars have masses $m\in(0.1,150)\,M_\odot$ and were drawn from the cluster's IMF up to a total added stellar mass of 10\% of the cluster mass. Each added star was assigned a random age between $1$ and $6\,\mathrm{Myr}$. We computed the corresponding $L_{\mathrm{H}\alpha}$  as described in Sect.~\ref{subsec:Pegase} and added it to the cluster total, together with the corresponding mass contribution.
We adopted the 10\% contamination level and the $1$--$6\,\mathrm{Myr}$ age range as a conservative upper-limit bracketing case, intended to test how strongly residual contamination could affect the inferred cluster distributions.

\subsection{Mean and standard deviation}\label{sec:Meth_MeanStd}
To quantify the results better, the weighted mean is calculated with Eq.~(\ref{eq:mean}) as J23 did. The standard deviation of the mean is determined with Eq.~(\ref{eq:Std}) from \citet{Taylor_Stdv}, where $N$ describes the number of star clusters:\begin{align}
        \Biggl \langle \frac{L_{\mathrm{H}\alpha}}{M_\mathrm{cl}} \Biggr \rangle &= \log_{10} \Bigg(\frac{\sum_i L_{\mathrm{H}\alpha, i}}{\sum_i M_{\mathrm{cl}, i}} \Bigg ) \label{eq:mean}\\
        \sigma_{\langle L_{\mathrm{H}\alpha} /M_\mathrm{cl} \rangle} &= 
        \frac{1}{\sqrt{N}}
        \sqrt{\frac{1}{N}\sum_i \Bigg(\log_{10}\Big(\frac{L_{\mathrm{H}\alpha, i}}{M_{\mathrm{cl}, i}}\Big) - 
                \frac{1}{N}\sum_i \log_{10} \Big(\frac{L_{\mathrm{H}\alpha, i}}{M_{\mathrm{cl}, i}}\Big)\Bigg)^2.} \label{eq:Std}
\end{align}
J23 divided the clusters into three different mass ranges, namely $(5.17-23.17)\,\times10^2\,M_\odot$, $(2.35-14.52)\,\times10^3\,M_\odot$ and $(1.881-8.567)\,\times10^4\,M_\odot$. 
The clusters of this work were divided into similar bins but not exactly the same ones, namely $(5.0-22.9)\,\times10^2\,M_\odot$, $(2.3-14.9)\,\times10^3\,M_\odot$, and $(1.5-10.0)\times10^4\,M_\odot$ because of different cluster masses. The three mass ranges are called Bins 1, 2, and 3, respectively. 
In addition, the weighted mean and standard deviation were calculated for all clusters (Bin 5) of this work, so for masses in  $(316-10^5)\,M_\odot$ and for clusters that are in the range of the clusters observed by J23 (Bin 4). For a better comparison to J23,  the same number of clusters as in J23 is picked randomly from the simulated data and the mean is calculated for the respective bin. This is repeated 1000 times from which the median and the median absolute deviation (MAD) are determined. Note that throughout this work, we drop the units for conciseness, but solar units are always implied: $\langle L_{\mathrm{H}\alpha} /M_\mathrm{cl} \rangle =\Bigl \langle \frac{L_{\mathrm{H}\alpha} /M_\mathrm{cl}}{L_{\mathrm{H}\alpha, \odot} /M_\odot} \Bigr \rangle$. 

\section{Results}\label{sec:Results}
In Sects.~\ref{sec:Results_stellarevolution} and \ref{sec:Results_Halpha} the computational results of the stellar evolution and the H$\alpha$ luminosity are presented. In Sect.~\ref{sec:Results_sampled_clusters} the sampled clusters are depicted under different assumptions, such as dynamical ejections and observational uncertainties.

\subsection{Stellar evolution}\label{sec:Results_stellarevolution}
In Fig.~\ref{fig:Mnow_Mini} the stellar mass evolution of stars with ages of $1-4\,\mathrm{Myr}$ and different metallicities is displayed.  The stellar mass evolution for stars with ages of $5-7\,\mathrm{Myr}$ is displayed in Fig.~\ref{fig:Mnow_Mini_5_6}.
\begin{figure}[ht]
        \centering
        \includegraphics[width=0.45\textwidth]{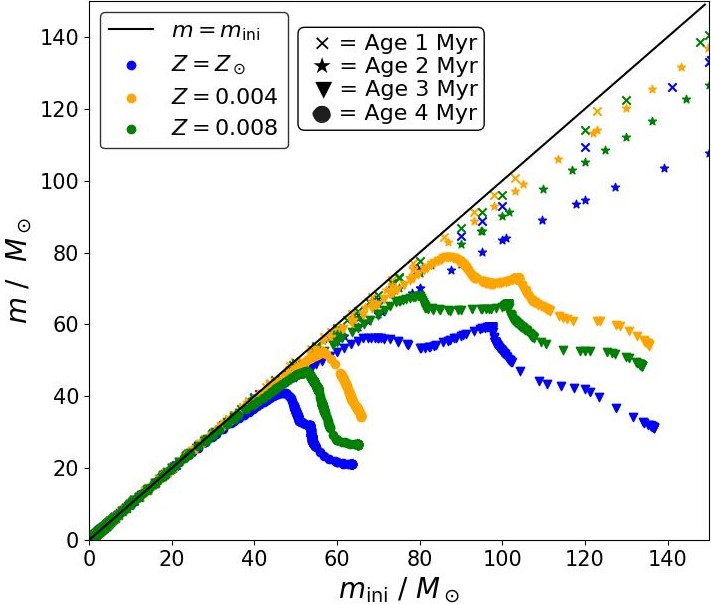}
        \caption{Stellar mass evolution for ages of $1-4\,\mathrm{Myr}$ obtained as described in Sect.~\ref{sec:stellar_evolution}. The mass of a star at the respective age, $m$, is plotted against the initial mass, $m_\mathrm{ini}$. The colour indicates the metallicity, and the marker the age. The black line displays the case $m = m_\mathrm{ini}$.}
        \label{fig:Mnow_Mini} 
\end{figure}

\noindent For low $m_\mathrm{ini}$, the mass at an age of $1\,\mathrm{Myr}$ does not change but the mass of stars with $m_\mathrm{ini} > 60\,M_\odot$ decreases. The higher the initial mass, the stronger the mass loss.
The stellar mass evolution is only displayed for $m_\mathrm{ini} \leq 150\,M_\odot$ because this is the upper mass limit of the IMF and no mergers leading to higher masses are included in this analysis.
At an age of $2\,\mathrm{Myr}$, stars with $m_\mathrm{ini}>40\,M_\odot$ undergo notable mass loss, especially stars with solar metallicity. 
At an age of $3\,\mathrm{Myr}$, no stars heavier than $m_\mathrm{ini}\approx 140\,M_\odot$ exist. Furthermore, the dependence between $m_\mathrm{ini}$ and $m$ becomes more complex.
At an age of $4\,\mathrm{Myr}$, stars up to a mass of $30\,M_\odot$ have not changed their mass  appreciably, while heavier stars can lose more than half of their initial mass. Stars with $m_\mathrm{ini} > 65\,M_\odot$ have lost their entire mass and are therefore no longer present at this age.
The higher the metallicity, the stronger the mass loss. For low-mass stars, the metallicity does not change the evolution significantly at the depicted ages.
  Due to mass loss, a star with a higher initial mass can become less massive than a star with a lower initial mass at a given age. For example, a star with $m_\mathrm{ini} \approx 140\,M_\odot$ can reach a similar mass at $3\,\mathrm{Myr}$ as a star with $m_\mathrm{ini} \approx 30\,M_\odot$.
 
 \begin{figure*}[ht]
        \sidecaption
        \includegraphics[width =12cm]{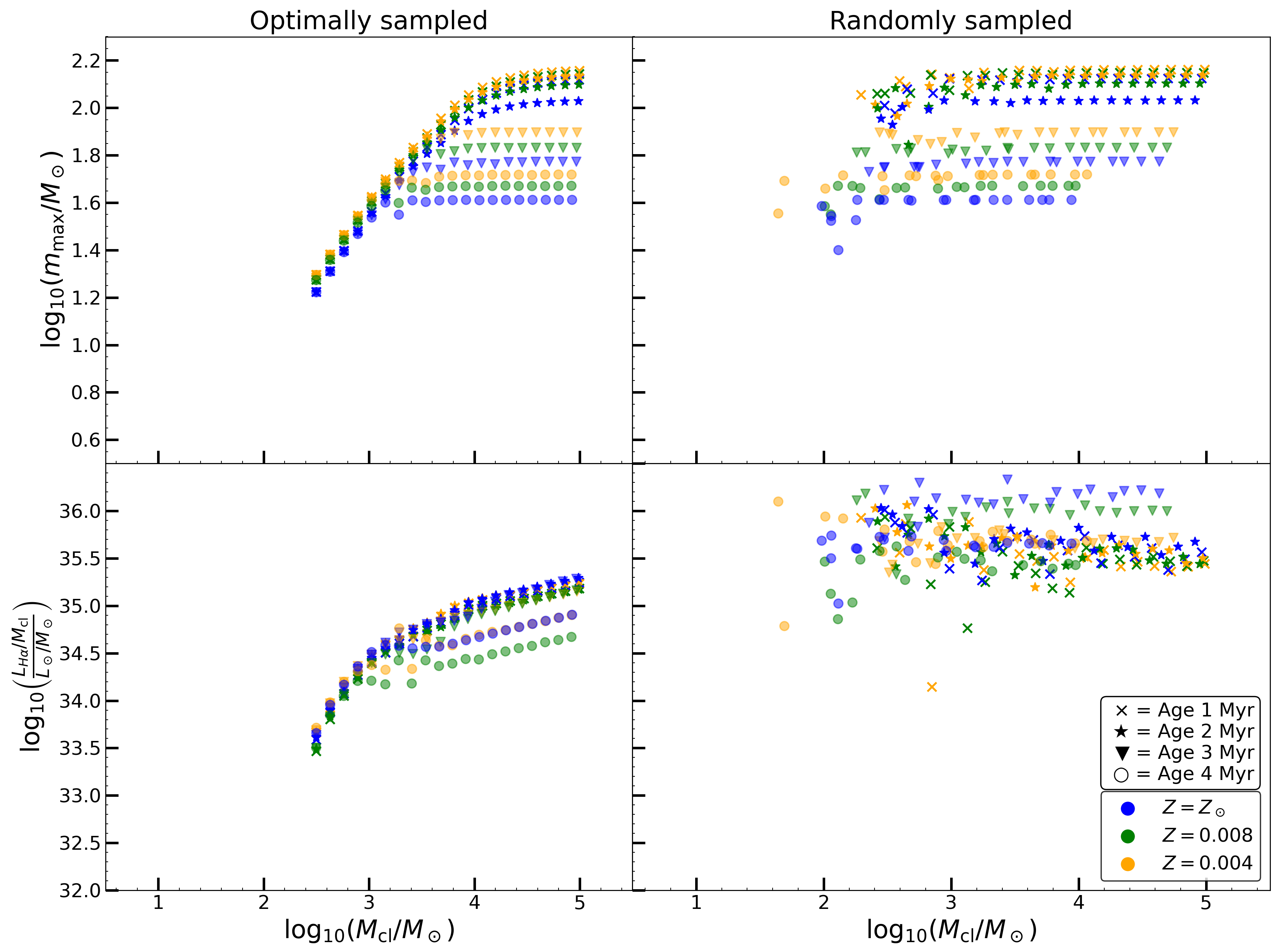}
        \caption{Optimally and randomly sampled clusters with ages of $1-4\,\mathrm{Myr}$. The colour indicates the metallicity of the cluster, and the marker the age. Top left: Mass of the most massive star in the cluster, $m_\mathrm{max}$, against the stellar mass of the cluster, $M_\mathrm{cl}$, for optimally sampled clusters. Top right: Same as  the top left but for randomly sampled clusters. Bottom left: Ratio of the cluster's H$\alpha$ luminosity and $M_\mathrm{cl}$ against $M_\mathrm{cl}$ for optimally sampled clusters. Bottom right: Same as the bottom left but for randomly sampled clusters.}
        \label{fig:Optimal_random_1-4_ini}
 \end{figure*}

\subsection{H$\alpha$ luminosity}\label{sec:Results_Halpha}
In Fig.~\ref{fig:LHalpha_Mini} the dependence of the H$\alpha$ luminosity, $L_{\mathrm{H}\alpha}$, on the initial stellar mass, $m_\mathrm{ini}$, is displayed for different metallicities and ages of $1-4\,\mathrm{Myr}$. These results are obtained with \texttt{P\'egase} as explained in Sect.~\ref{subsec:Pegase}.
The plots of the H$\alpha$ luminosity against the initial stellar mass   split by age ($1$--$7\,\mathrm{Myr}$) are shown in Fig.~\ref{fig:LHalpha_Mini_5_6}.

While in the higher-mass regime ($m>25\,M_\odot$), the curve is flat, for lower initial masses, $L_{\mathrm{H}\alpha}$ increases strongly. Thus, a different mass up to $m_\mathrm{ini} \approx 25\,M_\odot$ has a higher impact on $L_{\mathrm{H}\alpha}$ than a change of mass in the higher-mass regime. 
The data points for stars with an age of $3\,\mathrm{Myr}$ are mostly covered by the data points of stars with an age of $4\,\mathrm{Myr}$. At these ages, for $m_\mathrm{ini}>70\,M_\odot$, the values change without apparent correlation.
Since the model's upper mass limit is $120\,M_\odot$, $L_{\mathrm{H}\alpha}$ is extrapolated up to $m_\mathrm{max}=150\,M_\odot$ for ages at which these stars are still alive.  
There is no visible correlation between $L_{\mathrm{H}\alpha}$ and metallicity. 

\begin{figure}[ht]
        \centering
        \includegraphics[width=0.45\textwidth]{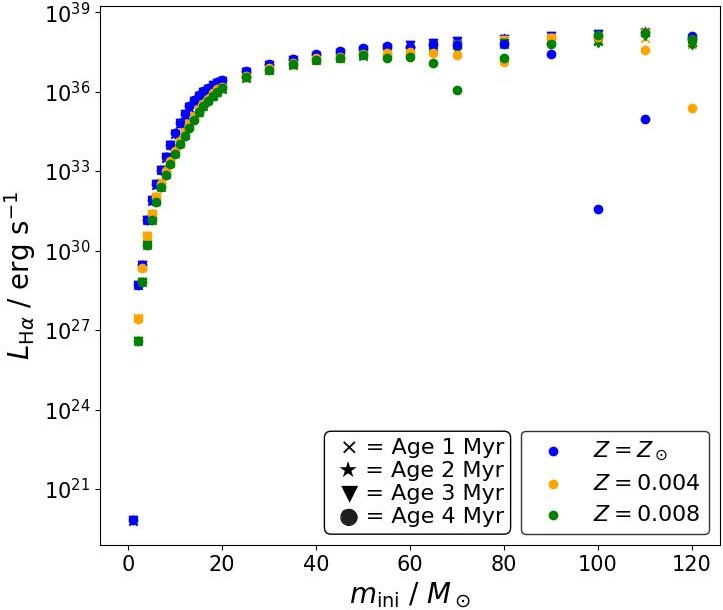}
        \caption{H$\alpha$ luminosity (logarithmic scale) of stars with ages of $1$--$4\,\mathrm{Myr}$ as a function of the initial mass, $m_\mathrm{ini}$. The colour indicates the metallicity, and the marker indicates the age. Note that the $1$--$3\,\mathrm{Myr}$ points are largely superposed by the $4\,\mathrm{Myr}$ data. In Appendix~\ref{sec:App_Halpha_1}, the data for each age are displayed in a separate figure for clarity.}
        \label{fig:LHalpha_Mini}
\end{figure}

\subsection{Sampled clusters}\label{sec:Results_sampled_clusters}
With the sampling methods explained in Sect.~\ref{sec:sampling_methods}, 160 star clusters with $M_\mathrm{ecl} = 10^{2.5}-10^5\,M_\odot$ and metallicities of $Z = \{0.014, 0.004, 0.008\}$ are sampled  {(see Sect.~\ref{sec:Methods} for details on the model cluster-mass selection and the matching procedure used for a fair comparison to J23).
} The resulting clusters, for which only the stellar evolution is accounted for, are displayed in Fig.~\ref{fig:Optimal_random_1-4_ini} and described in Sect.~\ref{subsec:initially}.
The clusters on which dynamical ejections are applied are shown in Sect.~\ref{subsec:Results_Dynamical_ejections}. The results in which the age uncertainty and the field stars are included are depicted in Sects.~\ref{sec:Res_age_uncertainty} and~\ref{sec:field_stars}, respectively.

In the upper panels of Figs.~\ref{fig:Optimal_random_1-4_ini} to~\ref{fig:Optimal_random_1-4_Der_1_MS3OP_max_2_1_Fs}, the mass of the most massive star, $m_\mathrm{max}$, is plotted against the mass of the stars in the cluster, $M_\mathrm{cl}$ at different ages. On the left, the optimally sampled clusters are displayed and on the right, the randomly sampled clusters.
In the lower panels, the ratio of the H$\alpha$ luminosity and $M_\mathrm{cl}$, $L_{\mathrm{H}\alpha}/M_\mathrm{cl}$, in dependence of $M_\mathrm{cl}$ is depicted.

\subsubsection{Including stellar evolution}\label{subsec:initially}
\begin{figure*}[ht]
\sidecaption
        \includegraphics[width =12cm]{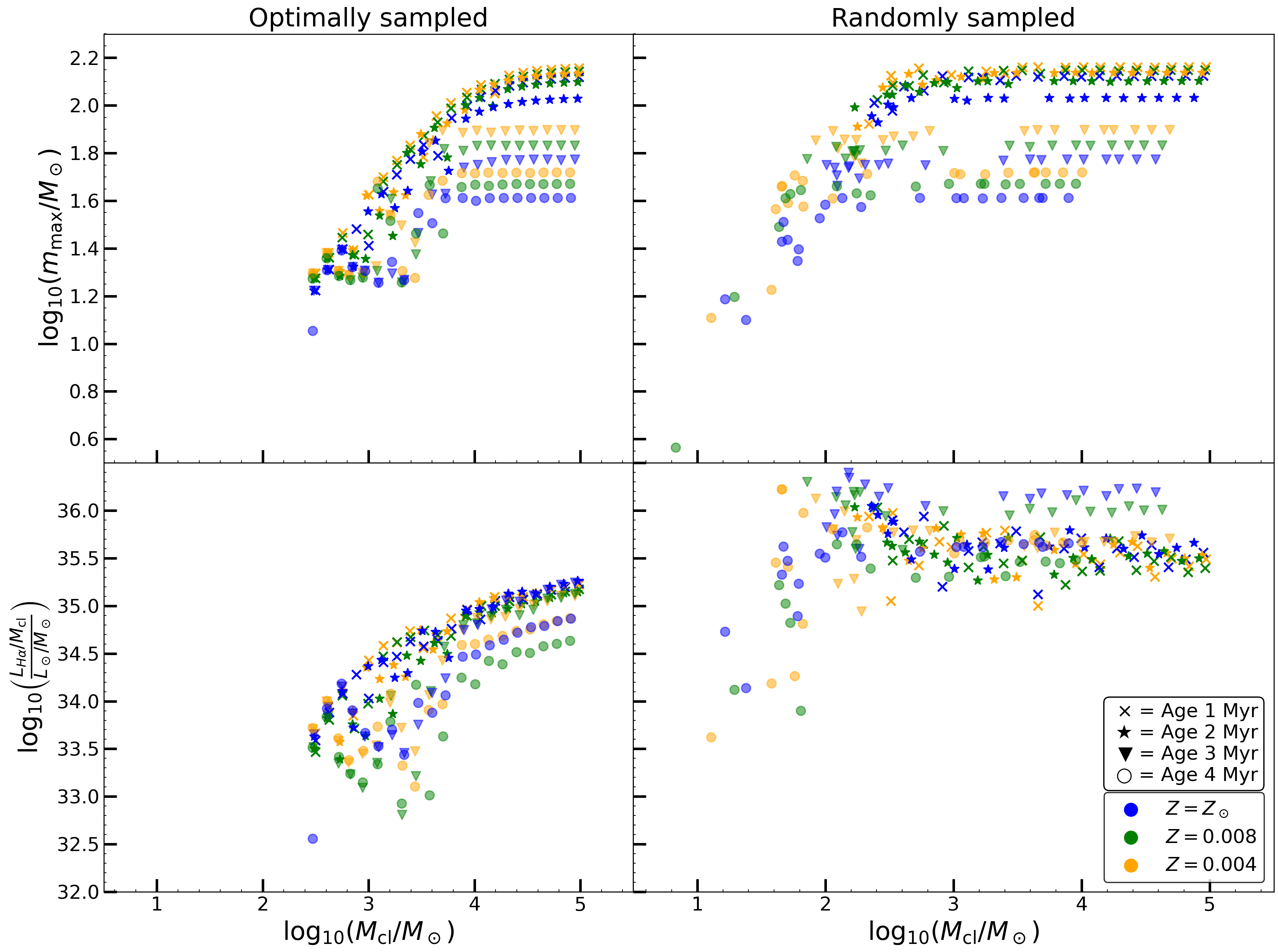}
                \caption{Same as Fig.~\ref{fig:Optimal_random_1-4_ini} but additionally including dynamical ejections with a $2\sigma$ higher ejection fraction.}
                \label{fig:Optimal_random_1-4_Der_1_MS3OP_max_2}
\end{figure*} 
For the optimally sampled clusters displayed in the upper left panel of Fig.~\ref{fig:Optimal_random_1-4_ini}, a clear correlation between $m_\mathrm{max}$ and $M_\mathrm{cl}$, the $m_\mathrm{max}-M_\mathrm{ecl}$ relation, is visible. Different plotted ages lead to changing break points at which the relation flattens. During a cluster's evolution, the stars develop and experience mass loss, which increases with initial stellar mass (see Fig.~\ref{fig:Mnow_Mini}). So, when the cluster ages, the mass of the most massive star decreases, which results in the lower break points. For the oldest clusters, some deviations from the $m_\mathrm{max}-M_\mathrm{ecl}$ relation are visible. This is caused by the further evolution of the stars as displayed in Fig.~\ref{fig:Mnow_Mini}.

A lower metallicity results in a more massive $m_\mathrm{max}$. This is a result of the varying IMF becoming top-heavier with lower metallicities (see Sect.~\ref{sec:meth_IMF}). 

The randomly sampled clusters do not show a clear relation between $m_\mathrm{max}$ and $M_\mathrm{cl}$. However, the samples exhibit plateaus that vary with age and metallicity due to the upper-mass limit $m_\mathrm{up}=150\,M_\odot$.
Some clusters are missing or do not follow the respective plateau, especially clusters with $M_\mathrm{cl}<1000\,M_\odot$. Fewer of these clusters host a star with a mass close to the upper mass limit because of the lower total mass. 
Furthermore, the randomly sampled clusters are able to host heavier stars and a lower number of objects than the optimally sampled clusters. 
After the death of the heaviest star, the second most massive star is the new most massive star whose mass is different to the original $m_\mathrm{max}$, which leads to a plateau.
An important aspect is that clusters with an age of $4\,\mathrm{Myr}$ are  shifted towards lower $M_\mathrm{cl}$. In the random-sampling case, some clusters can consist of only a few very massive stars; once these stars evolve off, the cluster loses a substantial fraction of its stellar mass and shifts to lower $M_\mathrm{cl}$.

In the lower panels, we plot the ratio $L_{\mathrm{H}\alpha}/M_\mathrm{cl}$ as a function of $M_\mathrm{cl}$. For the optimally sampled clusters, $L_{\mathrm{H}\alpha}/M_\mathrm{cl}$ increases with increasing $M_\mathrm{cl}$, indicating that the ionising output grows faster than the stellar mass. This is expected because clusters with $M_\mathrm{cl}\gtrsim 500\,M_\odot$ typically contain stars more massive than $\approx20\,M_\odot$, for which $L_{\mathrm{H}\alpha}$ is  orders of magnitude higher than for lower-mass stars. The slope of the   {$L_{\mathrm{H}\alpha}/M_\mathrm{cl}$--$M_\mathrm{cl}$ relation decreases towards higher cluster masses, consistent with Fig.~\ref{fig:LHalpha_Mini}, because the H$\alpha$ luminosity of individual stars changes only weakly for $m_\mathrm{ini}\gtrsim 25\,M_\odot$ in our models. Even once $m_\mathrm{max}$ approaches the upper-mass limit, $L_{\mathrm{H}\alpha}/M_\mathrm{cl}$ can still increase because additional ionising sources are added, rather than a single more luminous star. For clusters older than $\approx3\,\mathrm{Myr}$, stellar mass loss and  the more complex mapping between $L_{\mathrm{H}\alpha}$ and $m_\mathrm{ini}$ weaken the correlation. Finally, we do not find a strong systematic metallicity trend in $L_{\mathrm{H}\alpha}/M_\mathrm{cl}$ over the range considered here.

The ratio $L_{\mathrm{H}\alpha}/M_\mathrm{cl}$ of the randomly sampled clusters does not show an apparent correlation to $M_\mathrm{cl}$. Here, no gradient is visible compared to the optimally sampled clusters.

In J23, most clusters lie in the first bin (see Table~\ref{tab:Mean_Stdv_ini}). In our model set,  the optimally and randomly sampled clusters are distributed more uniformly across the first three bins. The fourth bin indicates that not all sampled clusters fall within the mass range analysed by J23, because our initial model grid spans a wider range to follow the evolution with time.
The medians of the weighted means of the randomly sampled clusters have a range of $\langle L_{\mathrm{H}\alpha} /M_\mathrm{cl} \rangle = (35.55-35.65)$ in the first three bins and the optimally sampled clusters of $\langle L_{\mathrm{H}\alpha} /M_\mathrm{cl} \rangle = (34.31-35.23)$. The values of the weighted mean in J23 range from $\langle L_{\mathrm{H}\alpha} /M_\mathrm{cl} \rangle = (33.900-34.176),$ which are lower than the results from both sampling methods. 

\subsubsection{Including dynamical ejections}\label{subsec:Results_Dynamical_ejections}
\begin{figure*}[ht]
\sidecaption
        \includegraphics[width=12cm]{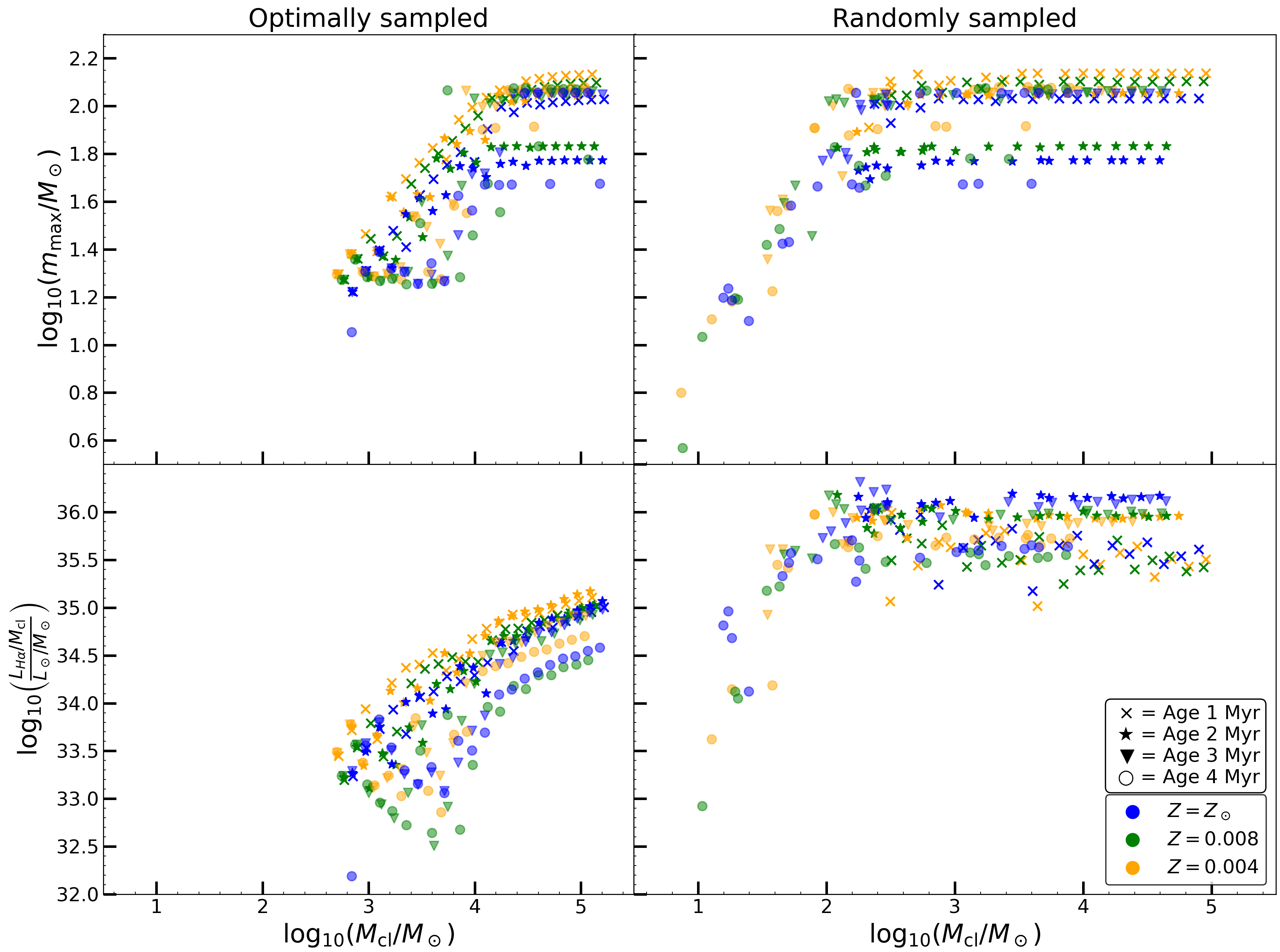} 
                \caption{Same as Fig.~\ref{fig:Optimal_random_1-4_ini} but including dynamical ejections with a $2\sigma$ higher ejection fraction and an age uncertainty of 1 Myr.}
                \label{fig:Optimal_random_1-4_Der_1_MS3OP_max_2_1}
\end{figure*}
The results for an ejection fraction within $2\sigma$ confidence are plotted in Fig.~\ref{fig:Optimal_random_1-4_Der_1_MS3OP_max_2}.
  With the higher ejection fraction, the $m_\mathrm{max}$--$M_\mathrm{ecl}$ relation for the optimally sampled clusters becomes more dispersed. For clusters with $M_\mathrm{cl} \lesssim 1500\,M_\odot$ it is no longer clearly visible as a tight relation, but remains as an upper envelope.
The ejection of the most massive stars reduces $m_\mathrm{max}$, while $M_\mathrm{cl}$ changes less because most of the cluster mass resides in low-mass stars that are unlikely to be ejected. As a result, the clusters move downwards and slightly leftwards in the diagram, but do not show the extreme shifts towards very low $M_\mathrm{cl}$ seen in some randomly sampled cases. The high-mass clusters remain largely unchanged because the adopted ejection-fraction model is limited to (and decreases towards) the highest masses covered by the simulations.
In the lower panel, the ratios of the optimally sampled clusters appear more scattered in the range where dynamical ejections are applied. Older clusters tend to show lower $L_{\mathrm{H}\alpha}/M_\mathrm{cl}$ than younger clusters. Since $M_\mathrm{cl}$ decreases, $L_{\mathrm{H}\alpha}$ must decrease even more strongly, consistent with the preferential ejection of the most massive (ionising) stars.

One cluster lies well below the main population (left panels at $\log_{10}(M_\mathrm{cl}/M_\odot) \approx 2.5$). The cluster has lost its most massive star but has not shifted markedly to lower $M_\mathrm{cl}$, implying that most of its remaining mass resides in low-mass stars. Since these stars contribute little to the ionising photon budget, the loss of a massive star reduces $L_{\mathrm{H}\alpha}$ substantially and therefore decreases $L_{\mathrm{H}\alpha}/M_\mathrm{cl}$.

In the upper-right panel of Fig.~\ref{fig:Optimal_random_1-4_Der_1_MS3OP_max_2}, where the randomly sampled clusters are displayed, a gap appears at $\log_{10}(M_\mathrm{cl}/M_\odot) \approx 3.0$ after applying ejections. This can occur because some randomly sampled clusters contain a small number of massive stars and comparatively few low-mass stars; when the massive stars are ejected, the remaining stellar mass drops sharply and the cluster shifts leftwards in the diagram.
This effect leads to a correlated looking population of clusters for $M_\mathrm{cl} \approx (10-100)\,M_\odot$. 
  Note that the clusters populating the lower-left corner consist of only one or a few remaining stars. Furthermore, the randomly sampled clusters are overall shifted further to the left because of the strong mass loss of the most massive stars and the loss of stars due to dynamical ejections.

In the lower-right panel, the stars are not as dispersed as in the previous plots and are mostly populated in the range $\log_{10}(L_{\mathrm{H}\alpha}/M_\mathrm{cl}) \approx (35.0-36.5)$ except for the clusters that have lost a significant amount of their mass, which are populated lower and on the left.
Furthermore, clusters with masses of around $100\,M_\odot$ have higher values of $L_{\mathrm{H}\alpha}/M_\mathrm{cl}$ now. This indicates that $M_\mathrm{cl}$ has decreased significantly with some stars, which led to a high $L_{\mathrm{H}\alpha}$ still remaining in the cluster. 

With the higher ejection fraction, the randomly sampled clusters have lost a significant amount of mass and 20 clusters moved out of the mass range by J23 to lower masses (see Table~\ref{tab:Mean_Stdv_Der_max2}). The values of $\langle L_{\mathrm{H}\alpha} /M_\mathrm{cl} \rangle$ are a little bit lower than before but still not in agreement with J23. 

\subsubsection{Including an age uncertainty}\label{sec:Res_age_uncertainty}
\begin{figure*}[ht]
\sidecaption
        \includegraphics[width=12cm]{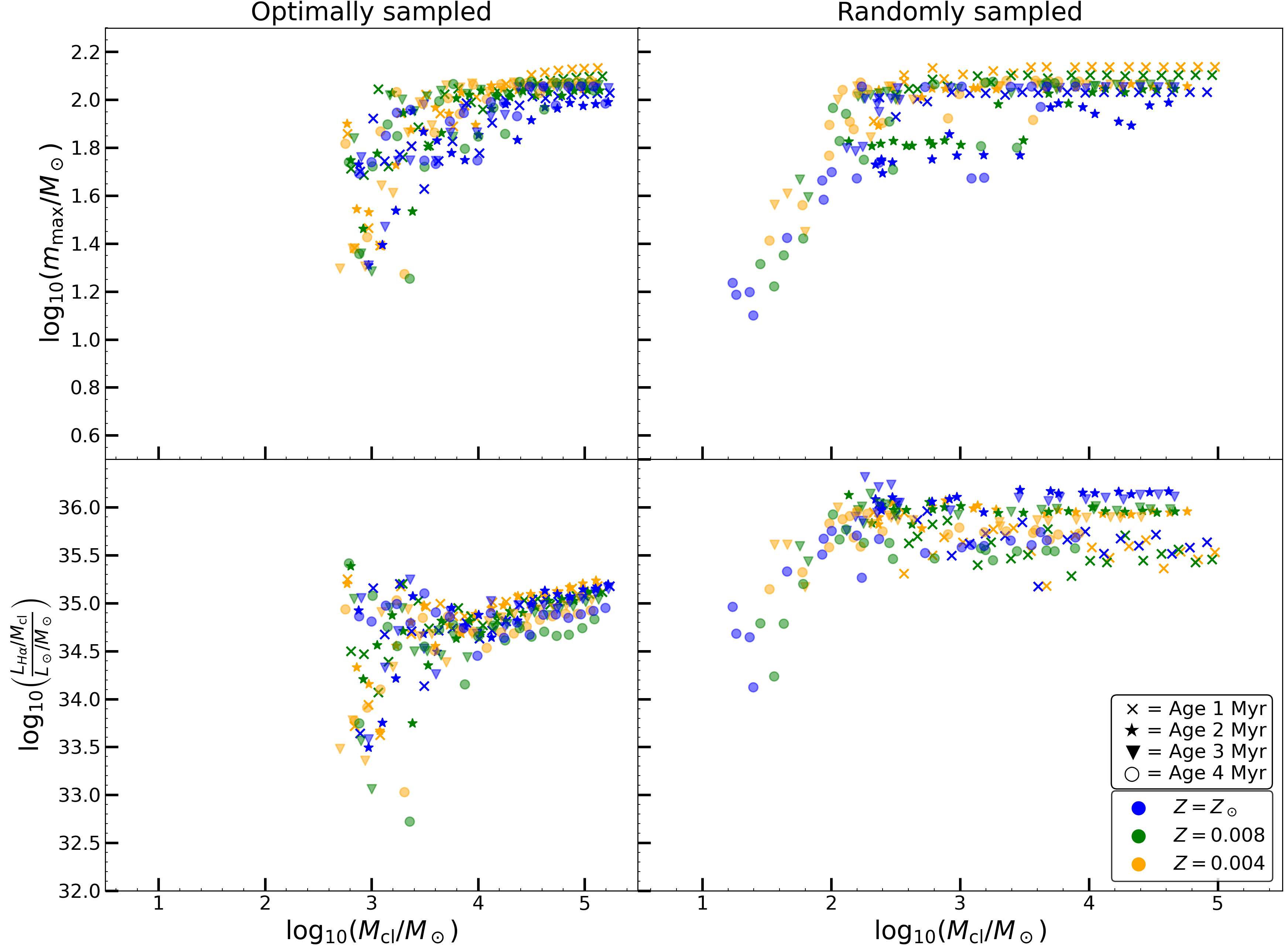} 
                \caption{Same as Fig.~\ref{fig:Optimal_random_1-4_ini} but including dynamical ejections with a $2\sigma$ higher ejection fraction, an age uncertainty of 1 Myr, and contamination of field stars.}
                \label{fig:Optimal_random_1-4_Der_1_MS3OP_max_2_1_Fs}
\end{figure*}
J23 question the age of the clusters and estimate that the clusters might be $1-2\,\mathrm{Myr}$ older. In our Figs.~\ref{fig:Optimal_random_1-4_Der_1_MS3OP_max_2_1}, \ref{fig:Optimal_random_1-4_Der_1_MS3OP_max_2_2}, and~\ref{fig:Optimal_random_1-4_Der_1_MS3OP_max_2_3}, the clusters are assumed to be $1\,\mathrm{Myr}$, $2\,\mathrm{Myr,}$ and $3\,\mathrm{Myr}$ older, respectively.

When we include an age uncertainty, the dispersion of the optimally sampled clusters increases in the upper-left panels. As the age uncertainty increases, the plateaus merge more strongly for both sampling methods. Older clusters tend to shift upwards because stellar masses are reassigned based on $L_{\mathrm{H}\alpha}$. The change in stellar mass increases with age (Figs.~\ref{fig:Mnow_Mini} and \ref{fig:Mnow_Mini_5_6}), and the maximum $L_{\mathrm{H}\alpha}$ produced by the $L_{\mathrm{H}\alpha}$--$m_\mathrm{ini}$ mapping decreases at older ages. As described in Sect.~\ref{sec:Meth_age_uncertainty}, if $L_{\mathrm{H}\alpha}$ exceeds the maximum value available at the shifted age in our precomputed grid, we used the youngest age bin considered for the $L_{\mathrm{H}\alpha}$--} $m_\mathrm{ini}$ mapping.

In the lower panel of  Fig.~\ref{fig:Optimal_random_1-4_Der_1_MS3OP_max_2_1}, the clusters also appear randomly populated except for the higher-mass clusters because of the lower dynamical ejection fractions for massive clusters. 
For larger assumed age uncertainties, the older star clusters move upwards, indicating that the inferred cluster mass decreases with increasing age uncertainty in this prescription.

The randomly sampled clusters show plateaus as before, but some clusters at ages of 3--4 Myr populate the upper plateau. This behaviour arises from the age-uncertainty prescription, which reassigns stellar masses based on a shifted-age $L_{\mathrm{H}\alpha}$--$m_\mathrm{ini}$ mapping (Sect.~\ref{sec:Meth_age_uncertainty}). Furthermore, about 25 low-mass clusters develop an apparent $m_\mathrm{max}$ {--} $M_\mathrm{cl}$ correlation under these assumptions.

In the lower-right panel, the clusters are mostly populated at values between $\log_{10}(L_{\mathrm{H}\alpha}/M_\mathrm{cl}) \approx 35-36$ except some older clusters that build a similar correlation like in the upper panel. Also, the spread of the randomly sampled clusters in the lower panel does not change significantly with age uncertainty.  

Compared to before, the number of randomly sampled clusters in the second and third bin decreased. On the other hand, the total number of randomly sampled clusters has not changed much, which implies that the clusters moved out of the mass range set in J23. Furthermore, the weighted mean of the randomly sampled clusters increased while the one of the optimally sampled clusters decreased as displayed in Table~\ref{tab:Mean_Stdv_Der_max_2_1}.
Since higher age uncertainties do not change the lower panels except for the massive optimally sampled clusters, which are not included in the set of clusters in J23, an age uncertainty of $1\,\mathrm{Myr}$ is assumed in the following plots.

\subsubsection{Including field stars}\label{sec:field_stars}

As described in Sect.~\ref{sec:Meth_fieldstars}, stars may be   {mis-assigned to clusters}.  {We accounted for this uncertainty by adding contaminating stars drawn from the same IMF as the cluster (Sect.~\ref{sec:Meth_fieldstars}).
} 

  {With added field-star contamination, the plateaus become less distinct}, as shown in the upper right panel of Fig.~\ref{fig:Optimal_random_1-4_Der_1_MS3OP_max_2_1_Fs}. Furthermore, all clusters shift upwards. Because   {additional stars are added, some clusters can contain a new $m_\mathrm{max}$. In addition, $M_\mathrm{cl}$ increases, which shifts all clusters slightly to the right.

In the lower panel, clusters with $M_\mathrm{cl} \approx 1000\,M_\odot$ are more dispersed. Some are shifted significantly upwards because of a new most massive star that leads to a higher $L_{\mathrm{H}\alpha}$ whose increase must exceed the one of $M_\mathrm{cl}$ because of the higher ratio $L_{\mathrm{H}\alpha}/M_\mathrm{cl}$. 
  High-mass clusters remain populated mostly in the same region, consistent with the upper panel. This is because high-mass clusters already host a larger number of massive stars than low-mass clusters, so adding a small contaminating component typically does not change their configuration as strongly.

The randomly sampled clusters change less under this prescription. Since they are already stochastically drawn, adding additional IMF-drawn stars primarily increases $M_\mathrm{cl}$ and produces a modest shift towards higher $M_\mathrm{cl}$.

  Referring to Table~\ref{tab:Mean_Stdv_Der_max_2_1_Fs}, adding field stars does not substantially change the resulting number of star clusters, weighted mean, and standard deviation for the randomly sampled clusters. For the optimally sampled clusters, $\langle L_{\mathrm{H}\alpha}/M_\mathrm{cl} \rangle$ increases compared to the case without field stars, consistent with Fig.~\ref{fig:Optimal_random_1-4_Der_1_MS3OP_max_2_1_Fs}.

\section{Discussion}\label{sec:Discussion}
In this section the results from Sect.~\ref{sec:Results} are discussed. First the influence of field stars and dynamical ejections are investigated. In Sect.~\ref{sec:final_comparison} the final plot (Fig.~\ref{fig:Last comparison}) and values are compared to the results by J23 and a $\chi^2$ test is applied in Sect.~\ref{sec:Disc_chi2}.

\subsection{Dynamical ejections and field stars} \label{sec:Disc_De_field_stars}
In Fig.~\ref{fig:Optimal_random_1-4_Der_1_MS3OP_mean_1_1_Fs} the clusters are displayed with dynamical ejections with a mean ejection fraction, an age uncertainty of $1\,\mathrm{Myr,}$ and randomly added field stars. Compared to Fig.~\ref{fig:Optimal_random_1-4_Der_1_MS3OP_max_2_1_Fs}, clusters with $M_\mathrm{cl}\approx 1000\,M_\odot$ are spread along the y-axis. On the lower right ($ \log_{10}(M_\mathrm{cl}/M_\odot) > 4$), this distribution is missing caused by the field stars. Thus, dynamical ejections move the star clusters downwards because of the ejections of massive stars, which reduce $L_{\mathrm{H}\alpha}$ more than $M_\mathrm{cl}$. Adding field stars moves the star clusters upwards in the plot because of the higher ratio of massive stars in the star cluster than before. This affects optimally sampled star clusters with masses around $1000\,M_\odot$ the most.
Furthermore, the dynamical ejections lead to the population of randomly sampled clusters in the lower-left corner, which is reduced with the added field stars.

The criterion for a star to be ejected is its distance from the cluster centre to be greater than $3\times  r_\mathrm{h}$ at $3\,\mathrm{Myr}$, and its velocity to be larger than the escape velocity at its distance \citep{Oh_Dependency_DynamicalEjections}. Star clusters expand in their evolution up to ten times of their initial half-mass radius \citep{Banerjee_clusterexpansion}. Thus, the clusters of this paper would have $r_\mathrm{h}\approx 3\,\mathrm{pc}$ at an age of $3\,\mathrm{Myr}$. This corresponds to a minimum distance of $9\,\mathrm{pc}$. Furthermore, the ejected systems have generally high velocities, which make them to move away from the cluster fast \citep{Oh_Dependency_DynamicalEjections}. Thus, those stars would no longer be in the frame of the aperture. A last note is that $3\times  r_\mathrm{h}$ is only the minimum distance; the real distance of the ejected systems might be greater.

\subsection{Final comparison}\label{sec:final_comparison}
  {Because the level of residual contamination and mis-assigned stars is uncertain, we treated the `field star' prescription as a bracketing case. In Fig.~\ref{fig:Last comparison} we therefore combine the model realisations with and without added field stars to illustrate the envelope of plausible outcomes once dynamical ejections are included. The corresponding summary statistics are listed in Table~\ref{tab:Mean_Stdv_Final}. For the comparison to J23, we restricted the model to the mass range covered by the observed sample and down-sampled the highest-mass model clusters to reflect the scarcity of high-mass clusters in J23.}

\begin{figure*}[ht]
        \centering
        \includegraphics[width=\linewidth]{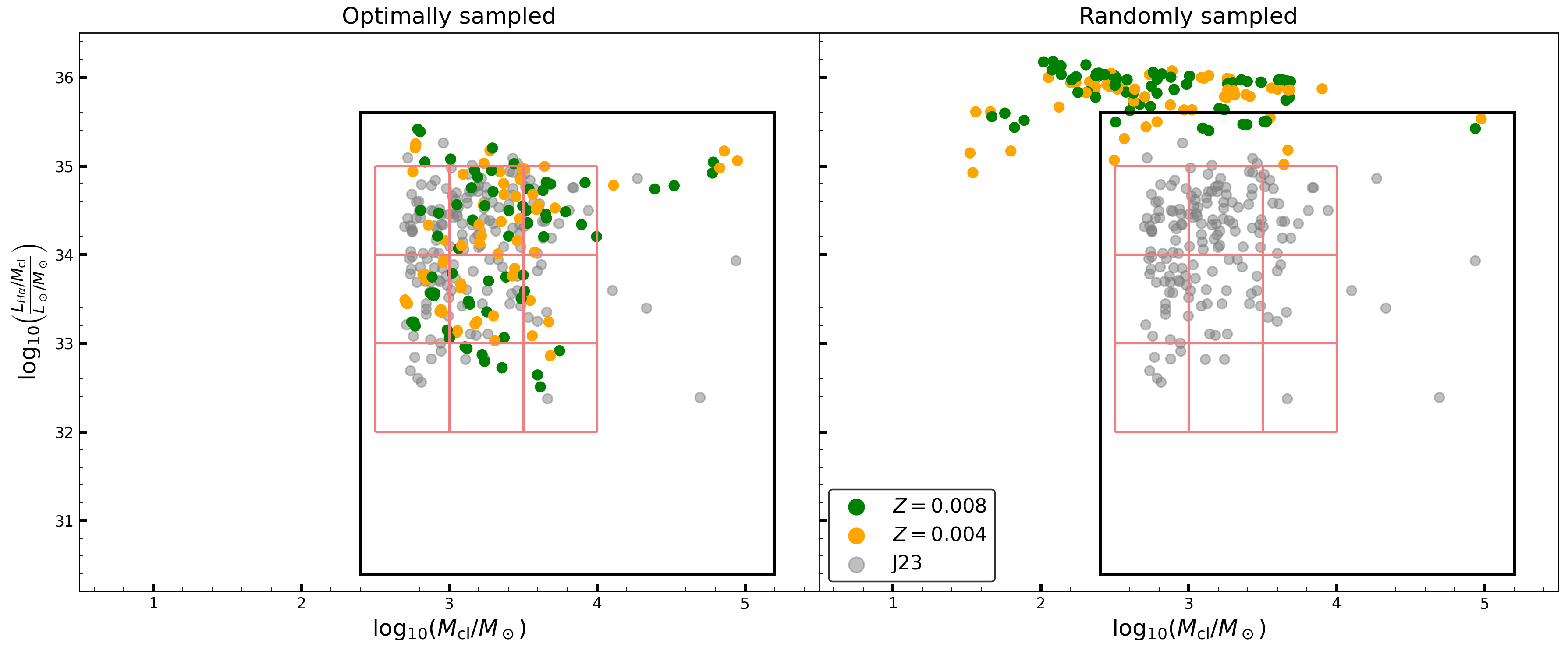}  
        \caption{Left: Sum of the lower-left panels of Figs.~\ref{fig:Optimal_random_1-4_Der_1_MS3OP_max_2_1} and~\ref{fig:Optimal_random_1-4_Der_1_MS3OP_max_2_1_Fs}. All ages are represented by the dots. The colour indicates the metallicity as before. The black box displays the limits  from J23, and the red grid shows the bins used for the $\chi^2$  test (Sect. \ref{sec:Disc_chi2}). Right: Same as the left but for randomly sampled clusters. The grey data points represent the observational data from J23 (see their Fig.~4 for  error bars, which are omitted here for clarity).}
        \label{fig:Last comparison}
\end{figure*}

To better quantify the results, the weighted mean and standard deviation are calculated as described in Sect.~\ref{sec:Meth_MeanStd}. The resulting values for clusters on which dynamical ejections with a $2\sigma$ higher ejection fraction, an age uncertainty of $1\,\mathrm{Myr}$ and on one set field stars and on the other no field stars are applied are listed in Table~\ref{tab:Mean_Stdv_Final}. 
The resulting weighted means of the optimally sampled clusters are in better agreement with J23 for all bins. The randomly sampled clusters have too high means. This is also reflected in Fig.~\ref{fig:Last comparison} where most of the randomly sampled clusters lie outside the range of J23.

The standard deviation provides a measure of dispersion in the data points. Except for the last bin, $\sigma_{\langle L_{\mathrm{H}\alpha} /M_\mathrm{cl} \rangle}$ for the optimally sampled clusters is about twice as high as for the randomly sampled clusters, indicating a larger dispersion. This is   consistent with Fig.~\ref{fig:Last comparison}, where the randomly sampled clusters  lie mainly in $\log_{10}( L_{\mathrm{H}\alpha}/M_\mathrm{cl}) \approx 35$--$36$ for $M_\mathrm{cl} > 100\,M_\odot$, while the optimally sampled clusters span approximately $\log_{10}(L_{\mathrm{H}\alpha}/M_\mathrm{cl}) \approx 32$--$35$.
In Bin 3,  $\sigma_{\langle L_{\mathrm{H}\alpha} /M_\mathrm{cl} \rangle}$ of the randomly sampled clusters is higher than for the optimally sampled clusters. That can be explained with the missing dynamical ejections in this range. However, these clusters do not influence the final comparison in Fig.~\ref{fig:Last comparison} because J23 do not observe a large number of massive star clusters. Therefore, the number of clusters is too low for a reliable comparison.
Furthermore, $\sigma_{\langle L_{\mathrm{H}\alpha} /M_\mathrm{cl} \rangle}$  cannot be compared in a strictly like-for-like way to the values reported by J23 because the exact estimator used there is not fully specified. If their dispersion measure is sensitive to sample size, then differences in the number of clusters per bin can also affect the reported values. We therefore interpret the dispersion comparison with caution and focus on the overall distributional agreement in the diagnostic plots.

Referring to Table~\ref{tab:Mean_Stdv_Final}, the number of optimally sampled clusters does not change in the respective mass bins significantly, which indicates that $M_\mathrm{cl}$ does not change with the added influences. 
On the other hand, the number of randomly sampled clusters decreases and $M_\mathrm{cl}$ shrinks because of the mass loss caused by the added influences.
Furthermore, the added effects lead to a higher $\sigma_{\langle L_{\mathrm{H}\alpha}/M_\mathrm{cl}\rangle}$ for the first four bins of the optimally sampled clusters while the values of the randomly sampled clusters decrease.

To facilitate a direct visual comparison to J23, we indicate the observed range from J23 with a black box in Fig.~\ref{fig:Last comparison}. We down-sampled the highest-mass model clusters so that the comparison is not dominated by cluster masses that are sparsely represented in the observed sample.

Referring to Fig.~\ref{fig:Last comparison}, the randomly sampled clusters are not in the range of the clusters analysed by J23, except for a few. Furthermore, in the cluster mass range, the ratios of $L_{\mathrm{H}\alpha}$ and $M_\mathrm{cl}$ do not vary that much as in J23.
For the optimally sampled clusters, displayed in the left panel, the clusters are spread similar to J23 and are in the same range as their results.
So, the optimally sampled clusters are more dispersed than the randomly sampled ones. 
This   {may appear counter-intuitive at first glance, but it illustrates that an initially ordered (deterministic) stellar population can evolve into a distribution that looks `stochastic' once evolutionary processes and observational uncertainties are taken into account}.

\subsection{$\chi^2$ test}\label{sec:Disc_chi2}
To compare the distributions further, we applied a $\chi^2$ test. The bin edges were chosen to be
$\log_{10}\Bigl(\frac{L_{\mathrm{H}\alpha}/M_\mathrm{cl}}{L_\odot/M_\odot}\Bigr) = 32, 33, 34, 35$ and $\log_{10}(M_\mathrm{cl}/M_\odot) = 2.5, 3.0, 3.5, 4.0$ and are displayed in Fig.~\ref{fig:Last comparison}. The number of data points for each bin cell was determined for the distribution in J23, $n_\mathrm{J,i}$, and for the calculated distribution, $n_\mathrm{c,i}$. The total number of data points is $N_\mathrm{J} = \sum_{i}^{N} n_\mathrm{J,i}$ for the data in J23 and $N_\mathrm{c} = \sum_{i}^{N} n_\mathrm{c,i}$ for the calculated data. In accordance with   {Eq}.~14.3.3 in \citet{Press_Chi2test}, the $\chi^2$ of binned data with an unequal number of data points was calculated as $\chi^2 = \sum_{i}^{N} \frac{(\sqrt{N_\mathrm{J}/N_\mathrm{c}}n_\mathrm{c,i}-\sqrt{N_\mathrm{c}/N_\mathrm{J}} n_\mathrm{J,i})^2}{n_\mathrm{c,i} + n_\mathrm{J,i}}$ and divided by the degrees of freedom, $\mathrm{d.o.f}=N_\mathrm{bin}-1$, where $N_\mathrm{bin}$ is the number of bins. For the optimally sampled clusters, $\frac{\chi^2}{\mathrm{d.o.f}} \approx 4$, and for the randomly sampled clusters, $\frac{\chi^2}{\mathrm{d.o.f}} \approx 17$. The corresponding p-values are $\approx10^{-4}$ and $\approx 10^{-25}$, which correspond to the $4\sigma$ and $11\sigma$ regions. This underlines the disagreement between the observed and randomly sampled clusters.

\subsection{Further comments}
There are additional effects, such as gas expulsion \citep{Kroupa_gasexpulsion, Dinnbier_gasexpulsion, Wirth_gasexpulsion}, which we did not include in this analysis. Gas expulsion can reduce the bound stellar mass and potentially introduce additional dispersion in the observable relations \citep{Weidner_infantweight}.
However, the optimally sampled models already show substantially better agreement with the observational distribution than the purely randomly sampled models under the set of effects explored here. In particular, the combination of stellar evolution, dynamical ejections, age uncertainty, and contamination can broaden the distribution of optimally sampled clusters in Fig.~\ref{fig:Last comparison}, whereas the randomly sampled clusters remain concentrated in a narrower region. We therefore conclude that evolutionary and observational effects can make an initially regulated stellar population appear more stochastic, and that a purely random sampling scenario does not reproduce the observed distribution within the framework investigated here.

\section{Conclusion}\label{sec:Conclusion}
In this study we analysed the evolution of star clusters using both optimal and random sampling methods. 
Clusters were modelled with masses between $10^{2.5}$ and $10^5\,M_\odot$ and metallicities of $Z = \{0.014, 0.004, 0.008\}$.
We examined how these sampling methods influence the $m_\mathrm{max}-M_\mathrm{ecl}$ relation and the ratio $L_{\mathrm{H}\alpha}/M_\mathrm{cl}$.

This study shows that, within the modelling framework explored here, optimal sampling yields distributions that are consistent with the extragalactic star cluster observations considered, while purely random sampling produces no agreement. Our modelling further suggests that the observed scatter in the LEGUS clusters can be accounted for by a combination of stellar evolution, dynamical ejections, age uncertainties, and residual contamination by field stars, rather than by the absence of the $m_{\text{max}} - M_{\text{ecl}}$ relation. 

For reproducibility, we provide our \texttt{Python}-based random-sampling implementation (linked in Sect.~\ref{sec:random_sampling}).
Overall, our results support a highly self-regulated interpretation of star formation and are consistent with key aspects of the integrated galaxy-wide IMF framework.

\begin{acknowledgements}
      We acknowledge support through the DAAD-EasternEurope Exchange grant at Bonn University and corresponding support from Charles University. PK acknowledges support through grant 26-217745 from the Czech Grant Agency.  {TJ acknowledges the support from the MUNI Award in Science and Humanities (MUNI/I/1762/2023).
} \end{acknowledgements}

\bibliography{Sources.bib}

\begin{appendix}
\section{H$\alpha$ luminosity}\label{sec:App_Halpha_1}
\noindent The downloaded files from \url{https://www2.iap.fr/users/fioc/PEGASE.html} comprise the stellar evolution tracks of the following metallicities \ $Z=\{0.0001, 0.0004, 0.004, 0.008, 0.02, 0.05, 0.1\}$.  
Furthermore, various IMFs are available, where the \citet{KTG_93} IMF is chosen as IMF input parameter.
\citet{KTG_93} describe the IMF as a three-part power law, where 
$\alpha_3 =  2.7$ for stars more massive than $1\,M_\odot$, $\alpha_2 = 2.2$ in the mass range $0.5\leq m/M_\odot \leq  1$ and $\alpha_1 = 1.3$ in the range $0.1 < m/M_\odot \leq 0.5$.
In addition, multiple filter and calibration options are available, which are not important for this work.

The first step in the execution is to create SSPs. As boundary masses, the masses $1\,M_\odot$ below and above the concerned mass are chosen. For example, if a \texttt{SSP} for $m_\mathrm{star} = 10\,M_\odot$ is set up, the lower boundary mass is $9\,M_\odot$ and the upper $11\,M_\odot$. For the supernovae ejecta, model B is selected, which contains the calculations of \citet{Woosley_modelB}. Furthermore, ejecta due to stellar winds in high-mass stars are accounted for.

After generating the \texttt{SSP}s, the \texttt{scenarios} are created. The fraction of close binary systems is set to 0 to allow normalisation to a single star later. The initial metallicity of the ISM is set to the respective metallicity of the galaxy or to the solar metallicity.
Furthermore, the galaxy is set to be already constituted and the star formation scenario to be an instantaneous burst $\mathrm{SFR} (t) = \delta (t)$. The fraction (in mass) of the star formation rate used to form substellar objects is set to 0. In addition, the galactic winds and the extinction are set to 0, while the nebular emission is accounted for.
After executing \texttt{spectra}, the normalised luminosities of the line with a wavelength of $\lambda = 6563\,\mathring {\mathrm A}$, which is similar to the wavelength of the H$\alpha$ line, are extracted.
\begin{figure*}[t]
        \centering
        \begin{subfigure}[t]{0.32\textwidth}
                \centering
                \includegraphics[width=\linewidth]{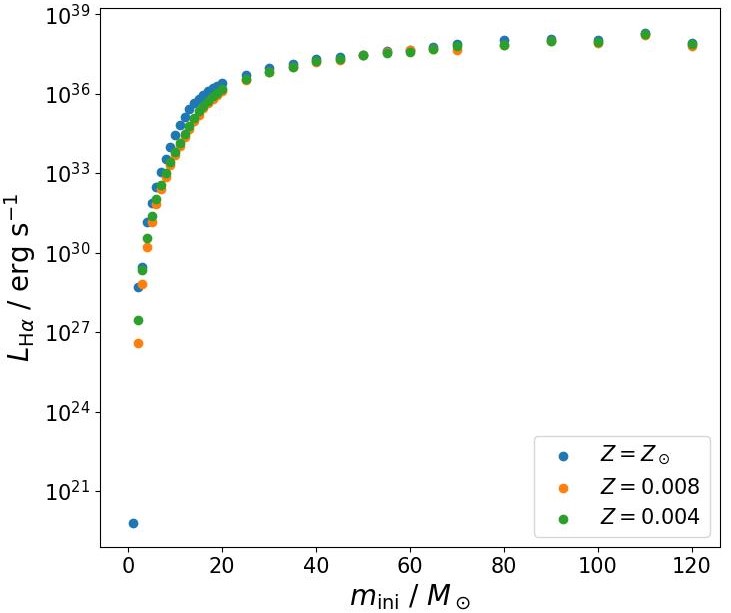}
        \end{subfigure}
        \begin{subfigure}[t]{0.32\textwidth}
                \centering
                \includegraphics[width=\linewidth]{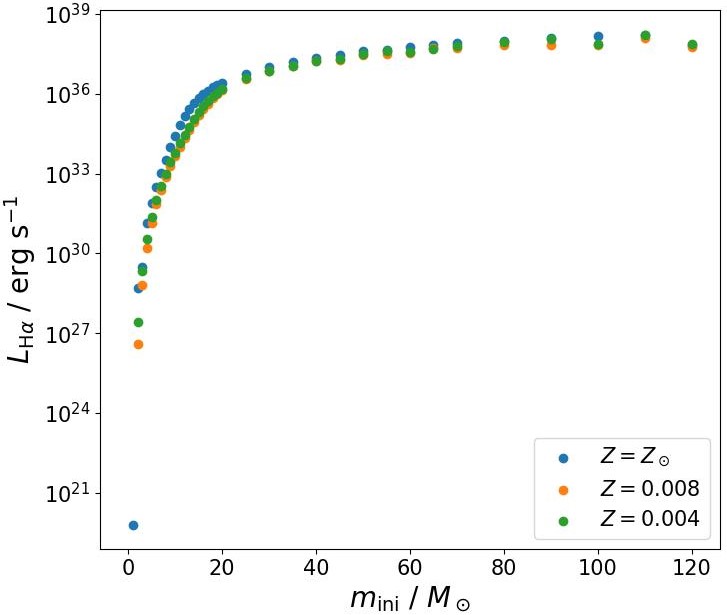}
        \end{subfigure}
        \begin{subfigure}[t]{0.32\textwidth}
                \centering
                \includegraphics[width=\linewidth]{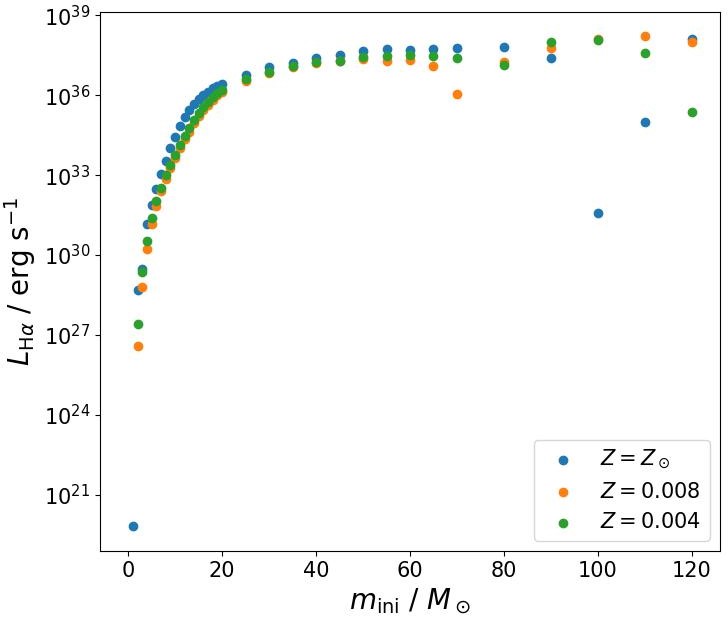}
        \end{subfigure}
        \vspace{0.3cm}
        \begin{subfigure}[t]{0.32\textwidth}
                \centering
                 \includegraphics[width=\linewidth]{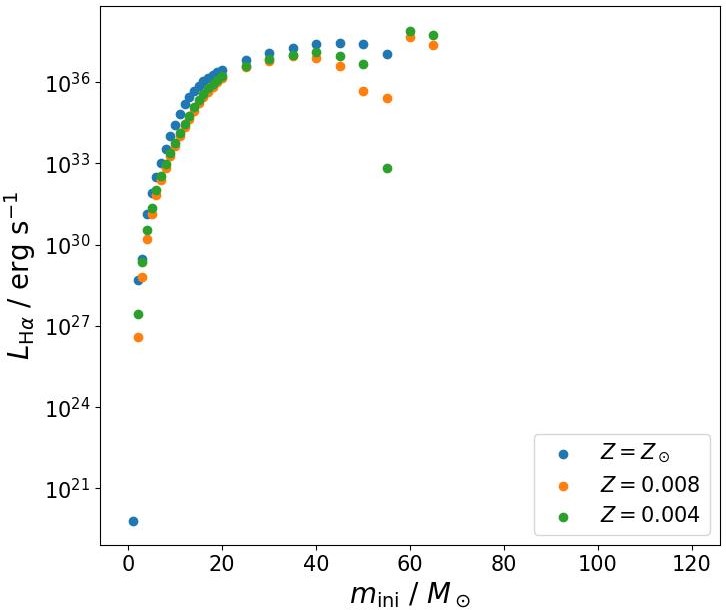}
        \end{subfigure}
        \begin{subfigure}[t]{0.32\textwidth}
                \centering
                 \includegraphics[width=\linewidth]{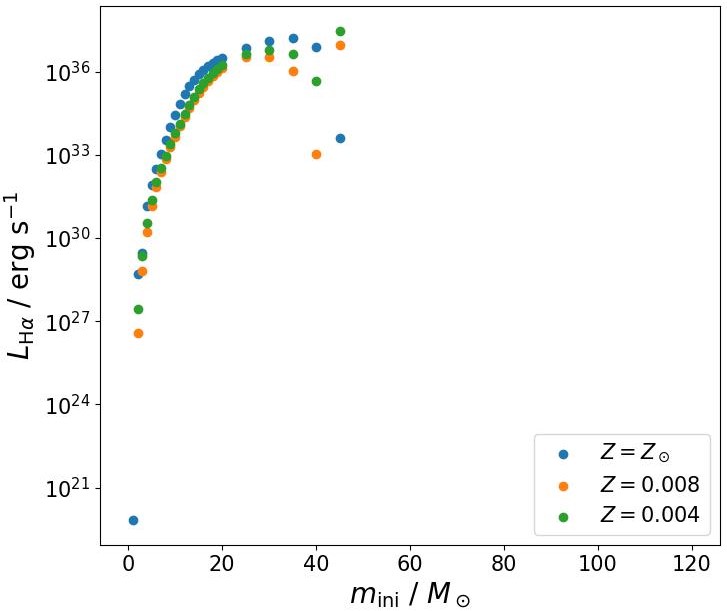}
        \end{subfigure}
        \begin{subfigure}[t]{0.32\textwidth}
                \centering
                \includegraphics[width=\linewidth]{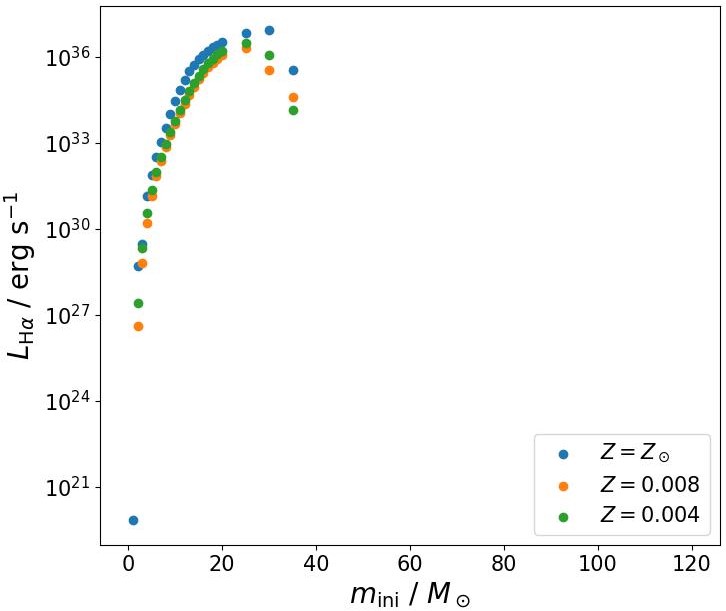}
        \end{subfigure}
        \vspace{0.3cm}
        \begin{subfigure}[t]{0.32\textwidth}
                \centering
                \includegraphics[width=\linewidth]{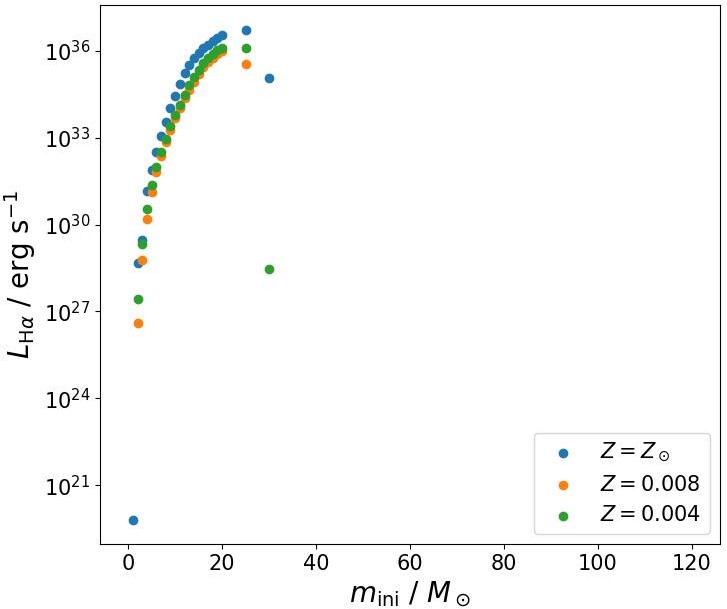}
        \end{subfigure}
        \caption{Same as Fig.~\ref{fig:LHalpha_Mini} but split by age for   $1$--$7\,\mathrm{Myr}$ (in order from $1\,\mathrm{Myr}$ to $7\,\mathrm{Myr}$).}
        \label{fig:LHalpha_Mini_5_6} 
\end{figure*}

\section{Dynamical ejections}
\begin{table}[H]
        \caption{Division of the spectral types and fraction conversion.}
        \label{tab:Ejection_Properties}
        \centering
        \begin{tabular}{c|c|c|c|c}
                &  O-type & early B-type & late B-type & A-type \\
                \hline
                $m/M_\odot$ & $ \geq 17.5$ & $5 \leq m
                < 17.5 $   & $3 < m \leq 5$ & $1.65 < m \leq 3$ \rule{0pt}{10pt}\\
                $f_\mathrm{conv}$ & 1 & 0.357 & 0.161 & 0.161
        \end{tabular}

\end{table}

\begin{table}[H]
        \caption{Values for the calculation of the dynamical ejections.}
        \label{tab:sigma_values}
        \centering
        \begin{tabular}{c|c|c|c|c}
                $M_\mathrm{ecl}$ & $\langle f_\mathrm{ej,o}\rangle$ &  $\langle f_\mathrm{b,ej,o}\rangle$ & $\sigma_{\langle f_\mathrm{ej,o}\rangle}$ & $N_\mathrm{run}$\\
                \hline
                $10^{2.5}$ & 0.080 & 0.78       & 0.08 & 100 \\
                $10^3$ & 0.172 & 0.33   & 0.18  & 100 \\
                $10^{3.5}$ & 0.252 & 0.17       & 0.25  & 100 \\
                $10^4$ & 0.149 & 0.13   & 0.07 & 10  \\
                $10^{4.5}$ & 0.105 & 0.05       & 0.05 & 4 \\
        \end{tabular}
        \tablefoot{In the first three columns, $M_\mathrm{ecl}$ and the corresponding $\langle f_\mathrm{ej,o}\rangle$ and $\langle f_\mathrm{b,ej,o}\rangle$ from \citet{Oh_Dependency_DynamicalEjections} are listed. The fourth column provides the values of $\sigma_{\langle f_\mathrm{ej,o}\rangle}$ that are added to $f_\mathrm{ej,O-sys}$ in order to increase the ejection fraction within sigma confidence regions. In the last column, the number of runs, $N_\mathrm{run}$, performed by \citet{Oh_Dependency_DynamicalEjections} are written.}

\end{table}

\section{Stellar evolution}\label{sec:App_Stellar_evolution}
\begin{figure}[H]
        \centering
        \begin{minipage}[t]{0.33\textwidth}
                \centering
                \includegraphics[width=\textwidth]{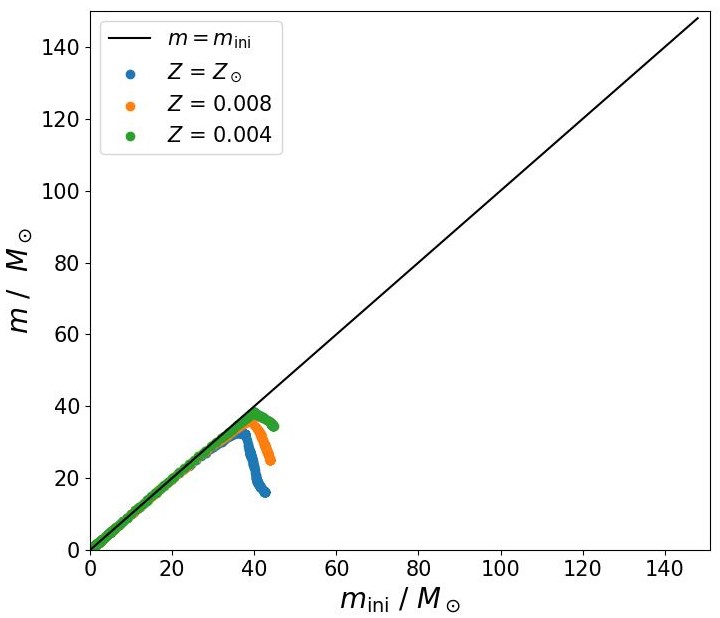}
        \end{minipage}
        \begin{minipage}[t]{0.33\textwidth}
                \centering
                \includegraphics[width=\textwidth]{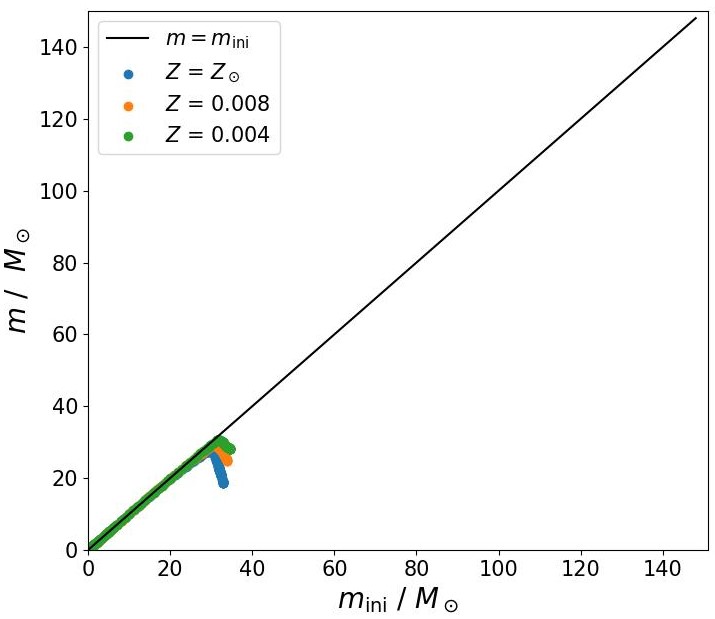}
        \end{minipage}
        \begin{minipage}[t]{0.33\textwidth}
                \centering
                \includegraphics[width=\textwidth]{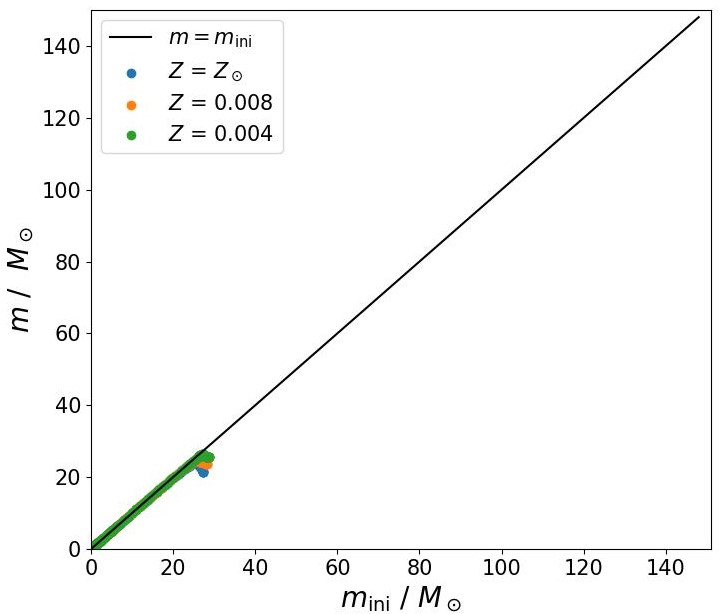}
        \end{minipage}
        \caption{Same as Fig.~\ref{fig:Mnow_Mini} but for ages of  $5\,\mathrm{Myr}$ (left), $6\,\mathrm{Myr}$ (middle), and $7\,\mathrm{Myr}$ (right).}
        \label{fig:Mnow_Mini_5_6}
\end{figure}

\section{Sampled clusters}\label{sec:App_Sampled_clusters}
\begin{table*}[t]
        \caption{Statistical results of the initial clusters.}
        \label{tab:Mean_Stdv_ini}
        \centering
        \begin{tabular}{c|c|c|c|c|c|c|c}
                Bin & Mass range ($M_\odot$) & Sampling method & $N$ & $\langle L_{\mathrm{H}\alpha} /M_\mathrm{cl} \rangle$ & $\sigma_{\langle L_{\mathrm{H}\alpha} /M_\mathrm{cl} \rangle}$ & Median & MAD\\
                \hline
                \multirow{3}{*}{1} & \multirow{2}{*}{$(5.0-22.9)\,\times10^2$} & optimal & 40 & $34.48$ & $0.03$ & $34.31$ & $0.01$ \rule{0pt}{10pt} \\
                &  & random & 38 & $35.67$ & $0.05$  & $35.55$ & $0.03$\\
                \cline{2-8}
                & $(5.17-23.17)\,\times10^2$ & J23 & 293 & $34.038$ & $0.016$ & &  \rule{0pt}{13pt} \\
                \hline
                \multirow{3}{*}{2} & \multirow{2}{*}{$(2.3-14.9)\,\times10^3$} & optimal &50 & 34.83 & 0.03  & $34.43$ & $0.01$ \rule{0pt}{10pt} \\
                &  & random & 48 & 35.67 & 0.03  & $35.65$ & $0.01$  \\
                \cline{2-8}
                & $(2.35-14.52)\,\times10^3$ & J23 & 73 & $34.176$ & $0.119$  &  & \rule{0pt}{13pt} \\
                \hline
                \multirow{3}{*}{3} & \multirow{2}{*}{$(1.5-10.0)\times10^4$} & optimal& 54 & 35.08 & 0.03  & $35.23$ & $0.1$ \rule{0pt}{10pt} \\
                & & random & 36 & 35.58 & 0.03  & $35.59$ & $0.01$ \\
                \cline{2-8}
                & $(1.881-8.567)\,\times10^4$ & J23 & 9 & $33.900$ & $0.079$  &  &  \rule{0pt}{13pt} \\
                \hline
                \multirow{2}{*}{4} & \multirow{2}{*}{$(0.5-100.0)\times10^3$} & optimal & 144 & 35.06 & 0.03  &  &  \rule{0pt}{10pt} \\
                & & random & 122 & 35.60 & 0.02  &  &  \\
                \hline
                \multirow{2}{*}{5} & \multirow{2}{*}{$316 - 10^5$} & optimal & 160 & 35.06 & 0.03  &  &  \rule{0pt}{10pt} \\
                & & random & 160 & 35.60 & 0.02   &  &  \\
        \end{tabular}
        \tablefoot{Number, weighted mean and standard deviation, median and median absolute deviation of the initial clusters are listed in the last five columns, respectively. The corresponding bin and sampling method are written in the first two columns. For comparison, the results by \citet{Jung_LEGUS} are listed. The bins are defined in Sect.~\ref{sec:Meth_MeanStd}.}
\end{table*}

\begin{table*}[t]
        \caption{Same as Table~\ref{tab:Mean_Stdv_ini} but with dynamical ejections with a mean ejection fraction.}
        \label{tab:Mean_Stdv_Der_mean}
        \centering
        \begin{tabular}{c|c|c|c|c|c|c|c}
                Bin & Mass range ($M_\odot$) & Sampling method & $N$ & $\langle L_{\mathrm{H}\alpha} /M_\mathrm{cl} \rangle$ & $\sigma_{\langle L_{\mathrm{H}\alpha} /M_\mathrm{cl} \rangle}$ & Median & MAD\\
                \hline
                \multirow{3}{*}{1} & \multirow{2}{*}{$(5.0-22.9)\,\times10^2$}& optimal & 40 & 34.42 & 0.03  & $34.077$ & $0.003$ \rule{0pt}{10pt}\\
                &  & random & 42 & 35.71 & 0.04  & $35.743$ & $0.003$ \\
                \cline{2-8}
                & $(5.17-23.17)\,\times10^2$ & J23 & 293 & $34.038$ & $0.016$  & &  \rule{0pt}{13pt} \\
                \hline
                \multirow{3}{*}{2} & \multirow{2}{*}{$(2.3-14.9)\,\times10^3$} & optimal & 51 & 34.79 & 0.03  & $34.743$ & $0.003$ \rule{0pt}{10pt} \\
                & & random &45 & 35.66 & 0.03  & $35.546$ & $0.004$  \\
                \cline{2-8}
                & $(2.35-14.52)\,\times10^3$ & J23 & 73 & $34.176$ & $0.119$ & & \rule{0pt}{13pt}   \\
                \hline
                \multirow{3}{*}{3} & \multirow{2}{*}{$(1.5-10.0)\times10^4$} & optimal& 53 & 35.08 & 0.03  & $34.998$ & $0.002$ \rule{0pt}{10pt} \\
                & & random & 34 & 35.59 & 0.03  & $35.5508$ & $0.0008$ \\
                \cline{2-8}
                & $(1.881-8.567)\,\times10^4$ & J23 & 9 & $33.900$ & $0.079$  &  & \rule{0pt}{13pt} \\
                \hline
                \multirow{2}{*}{4} & \multirow{2}{*}{$(0.5-100.0)\times10^3$} & optimal & 144 & 35.05 & 0.03  &  &  \rule{0pt}{10pt} \\
                & & random & 121 & 35.61 & 0.02  &  &  \\
                \hline
                \multirow{2}{*}{5} & \multirow{2}{*}{$316 - 10^5$} & optimal & 160 & 35.05 & 0.04  &  &  \rule{0pt}{10pt} \\
                & & random & 159 & 35.61 & 0.02  &  & \\
        \end{tabular}
\end{table*} 

\begin{table*}[t]
        \caption{Same as Table~\ref{tab:Mean_Stdv_ini} but with dynamical ejections with a $2\sigma$ ejection fraction.}
        \label{tab:Mean_Stdv_Der_max2}
        \centering
        \begin{tabular}{c|c|c|c|c|c|c|c}
                Bin & Mass range ($M_\odot$) & Sampling method & $N$ & $\langle L_{\mathrm{H}\alpha} /M_\mathrm{cl} \rangle$ & $\sigma_{\langle L_{\mathrm{H}\alpha} /M_\mathrm{cl} \rangle}$ & Median & MAD\\
                \hline
                \multirow{3}{*}{1} & \multirow{2}{*}{$(5.0-22.9)\,\times10^2$} & optimal & 45 & 34.06 & 0.07  & $34.428$ & $0.003$ \rule{0pt}{10pt} \\
                & &  random & 28 & 35.62 & 0.03  & $35.7601$ & $0.0001$ \\
                \cline{2-8}
                & $(5.17-23.17)\,\times10^2$ & J23 & 293 & $34.038$ & $0.0160$  &  &  \rule{0pt}{13pt} \\
                \hline
                \multirow{3}{*}{2} & \multirow{2}{*}{$(2.3-14.9)\,\times10^3$} & optimal & 48 & 34.75 & 0.07  & $34.894$ & $0.004$  \rule{0pt}{10pt}\\
                & & random & 41 & 35.64 & 0.04   & $35.368$ & $0.003$ \\
                \cline{2-8}
                & $(2.35-14.52)\,\times10^3$ & J23 & 73 & $34.176$ & $0.119$  &  &  \rule{0pt}{13pt} \\
                \hline
                \multirow{3}{*}{3} & \multirow{2}{*}{$(1.5-10.0)\times10^4$} &optimal& 51 & 35.06 & 0.03  & $35.076$ & $0.005$ \rule{0pt}{10pt} \\
                & &  random & 32 & 35.59 & 0.03  & $35.374$ & $0.004$ \\
                \cline{2-8}
                & $(1.881-8.567)\,\times10^4$ & J23 & 9 & $33.900$ & $0.0786$  &  &  \rule{0pt}{13pt} \\
                \hline
                \multirow{2}{*}{4} & \multirow{2}{*}{$(0.5-100.0)\times10^3$} & optimal & 144 & 35.02 & 0.05  &  &   \rule{0pt}{10pt}\\
                & & random & 101 & 35.6 & 0.02  &  & \\
                \hline
                \multirow{2}{*}{5} & \multirow{2}{*}{$316 - 10^5$} & optimal & 160 & 35.02 & 0.05  & &  \rule{0pt}{10pt} \\
                & &     random & 154 & 35.6 & 0.05  & &  \\
        \end{tabular}
\end{table*}

\begin{table*}[t]
        \caption{Same as Table~\ref{tab:Mean_Stdv_ini} but with dynamical ejections with a $2\sigma$ ejection fraction and an age uncertainty of $1\,\mathrm{Myr}$.}
        \label{tab:Mean_Stdv_Der_max_2_1}
        \centering
        \begin{tabular}{c|c|c|c|c|c|c|c}
                Bin & Mass range ($M_\odot$) & Sampling method & $N$ & $\langle L_{\mathrm{H}\alpha} /M_\mathrm{cl} \rangle$ & $\sigma_{\langle L_{\mathrm{H}\alpha} /M_\mathrm{cl} \rangle}$ & Median & MAD\\
                \hline
                \multirow{3}{*}{1} & \multirow{2}{*}{$(5.0-22.9)\,\times10^2$} &  optimal &  45 & 33.66 & 0.05  & $33.62$ & $0.003$ \rule{0pt}{10pt}\\
                & & random & 30 & 35.82 & 0.04  & $35.994$ & $0.006$\\
                \cline{2-8}
                & $(5.17-23.17)\,\times10^2$ & J23 & 293 & $34.038$ & $0.016$  &  &  \rule{0pt}{13pt} \\
                \hline
                \multirow{3}{*}{2} & \multirow{2}{*}{$(2.3-14.9)\,\times10^3$} & optimal & 51 & 34.35 & 0.08  & $33.843$ & $0.003$ \rule{0pt}{10pt}\\
                & &  random & 39 & 35.8 & 0.04   & $35.451$ & $0.002$ \\
                \cline{2-8}
                & $(2.35-14.52)\,\times10^3$ & J23 & 73 & $34.176$ & $0.119$ & & \rule{0pt}{13pt}   \\
                \hline
                \multirow{3}{*}{3} & \multirow{2}{*}{$(1.5-10.0)\times10^4$} & optimal& 54 & 34.8348 & 0.0352  & $34.941$ & $0.001$  \rule{0pt}{10pt}\\
                & & random & 27 & 35.75 & 0.05  & $35.3193$ & $0.0007$ \\
                \cline{2-8}
                & $(1.881-8.567)\,\times10^4$ & J23 & 9 & $33.900$ & $0.079$  &  &  \rule{0pt}{13pt} \\
                \hline
                \multirow{2}{*}{4} & \multirow{2}{*}{$(0.5-100.0)\times10^3$} & optimal & 150 & 34.79 & 0.06  &  &  \rule{0pt}{10pt}\\
                & & random & 96 & 35.76 & 0.02  &  &  \\
                \hline
                \multirow{2}{*}{5} & \multirow{2}{*}{$316 - 10^5$} & optimal & 160 & 34.85 & 0.05  &  &  \rule{0pt}{10pt} \\
                & & random & 153 & 35.77 & 0.06  &  &  \\
        \end{tabular}
\end{table*} 

\begin{table*}[t]
        \caption{Same as Table~\ref{tab:Mean_Stdv_ini} but with dynamical ejections with a $2\sigma$ higher ejection fraction, an age uncertainty of $1\,\mathrm{Myr,}$ and field stars.}
        \label{tab:Mean_Stdv_Der_max_2_1_Fs}
        \centering
        \begin{tabular}{c|c|c|c|c|c|c|c}
                Bin & Mass range ($M_\odot$) & Sampling method & $N$ & $\langle L_{\mathrm{H}\alpha} /M_\mathrm{cl} \rangle$ & $\sigma_{\langle L_{\mathrm{H}\alpha} /M_\mathrm{cl} \rangle}$ & Median & MAD \\
                \hline
                \multirow{3}{*}{1} & \multirow{2}{*}{$(5.0-22.9)\,\times10^2$} & optimal &  42 & 34.63 & 0.09  & $34.936$ & $0.005$ \rule{0pt}{10pt} \\
                & & random & 29 & 35.83 & 0.04 & $35.8$ & $0.2$ \\
                \cline{2-8}
                & $(5.17-23.17)\,\times10^2$ & J23 & 293 & $34.038$ & $0.016$  &  &  \rule{0pt}{13pt} \\
                \hline
                \multirow{3}{*}{2} & \multirow{2}{*}{$(2.3-14.9)\,\times10^3$} & optimal & 53 & 34.72 & 0.03  & $34.718$ & $0.002$ \rule{0pt}{10pt} \\
                & & random &40 & 35.8 & 0.03  & $35.983$ & $0.003$  \\
                \cline{2-8}
                & $(2.35-14.52)\,\times10^3$ & J23 & 73 & $34.176$ & $0.119$  &  &  \rule{0pt}{13pt} \\
                \hline
                \multirow{3}{*}{3} & \multirow{2}{*}{$(1.5-10.0)\times10^4$} & optimal & 53 & 34.98 & 0.02  & $35.032$ & $0.002$ \rule{0pt}{10pt} \\
                & & random & 27 & 35.76 & 0.04 & $35.925$ & $0.005$ \\
                \cline{2-8}
                & $(1.881-8.567)\,\times10^4$ & J23 & 9 & $33.900$ & $0.079$  &  &  \rule{0pt}{13pt} \\
                \hline
                \multirow{2}{*}{4} & \multirow{2}{*}{$(0.5-100.0)\times10^3$} & optimal & 148 & 34.95 & 0.03  &  &  \rule{0pt}{10pt} \\
                & & random & 96 & 35.77 & 0.02  &  &  \\
                \hline
                \multirow{2}{*}{5} & \multirow{2}{*}{$316 - 10^5$} & optimal & 160 & 35.0 & 0.03  &  &  \rule{0pt}{10pt} \\
                & & random & 153 & 35.77 & 0.03  &  &  \\
        \end{tabular}
\end{table*} 

\begin{table*}[t]
        \caption{Statistical results including all effects.}
        \label{tab:Mean_Stdv_Final}
        \centering
        \begin{tabular}{c|c|c|c|c|c|c|c}
                Bin & Mass range ($M_\odot$) & Sampling method & $N$ & $\langle L_{\mathrm{H}\alpha} /M_\mathrm{cl} \rangle$ & $\sigma_{\langle L_{\mathrm{H}\alpha} /M_\mathrm{cl} \rangle}$ & Median & MAD \\
                \hline
                \multirow{3}{*}{1} & \multirow{2}{*}{$(5.0-22.9)\,\times10^2$} &  optimal & 87 & $34.36$ & $0.07$  & $33.14$ & $0.06$ \rule{0pt}{10pt}\\
                &  & random & 59 & $35.825$ & $0.026$  & $35.4396$ & $0.0004$ \\
                \cline{2-8}
                & $(5.17-23.17)\,\times10^2$ & J23 & 293 & $34.038$ & $0.016$  & &  \rule{0pt}{13pt} \\
                \hline
                \multirow{3}{*}{2} & \multirow{2}{*}{$(2.3-14.9)\,\times10^3$} & optimal & 104 &$34.57$ & $0.06$ & $34.678$ & $0.003$ \rule{0pt}{10pt}\\
                & & random & 79 &$35.801$ & $0.026$  & $35.015$ & $0.005$ \\
                \cline{2-8}
                & $(2.35-14.52)\,\times10^3$ & J23 & 73 & $34.176$ & $0.119$  & &  \rule{0pt}{13pt} \\
                \hline
                \multirow{3}{*}{3} & \multirow{2}{*}{$(1.5-10.0)\times10^4$} & optimal& 107 & $34.913$ & $0.022$  & $34.976$ & $0.005$  \rule{0pt}{10pt}\\
                & &  random & 54 &$35.76$ & $0.03$  & $35.421$ & $0.001$ \\
                \cline{2-8}
                & $(1.881-8.567)\,\times10^4$ & J23 & 9 & $33.900$ & $0.079$  &  &  \rule{0pt}{13pt} \\
                \hline
                \multirow{2}{*}{4}  & \multirow{2}{*}{$(0.5-100.0)\times10^3$} & optimal & 298 &$34.88$ & $0.04$  &  &  \rule{0pt}{10pt}\\
                & & random & 192 & $35.767$ & $0.016$  &  &  \\
                \hline
                \multirow{2}{*}{5} & \multirow{2}{*}{$316 - 10^5$} & optimal & 320 &$34.94$ & $0.04$  &  &  \rule{0pt}{10pt}\\
                & & random & 306 & $35.77$ & $0.04$  &  &  \\
        \end{tabular}
        \tablefoot{Same as Table~\ref{tab:Mean_Stdv_ini} but for the summed clusters of cases with a $2\sigma$ ejection fraction, an age uncertainty of $1\,\mathrm{Myr, and}$ with and without field stars (see Sect.~\ref{sec:Disc_De_field_stars}) for the final comparison.}
\end{table*}

\begin{figure*}[t]
\sidecaption
        \includegraphics[width =12cm]{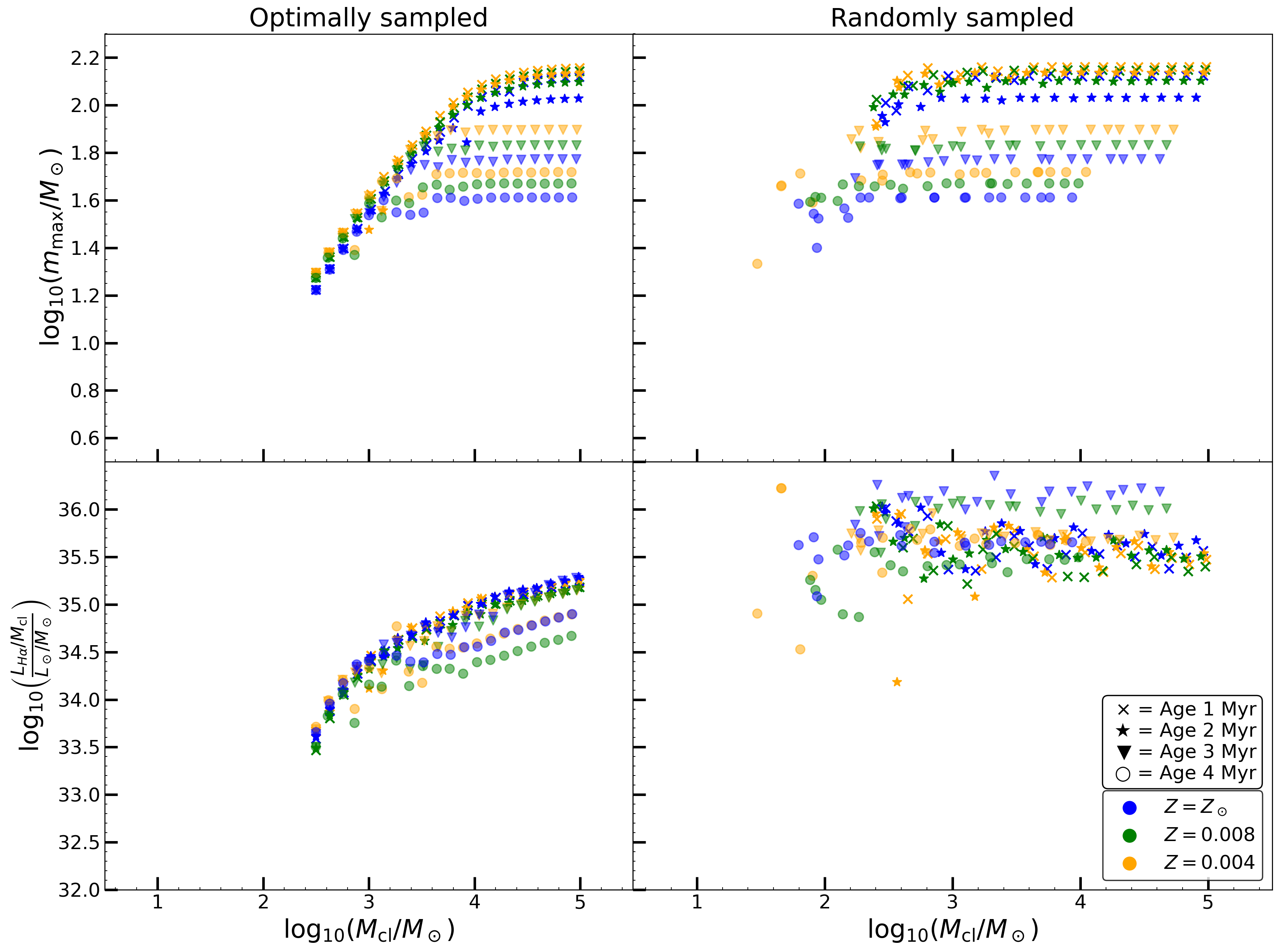}
                \caption{Same as Fig.~\ref{fig:Optimal_random_1-4_ini} but including dynamical ejections with the mean ejection fraction.}
                \label{fig:Optimal_random_1-4_Der_1_MS3OP_mean_1}
\end{figure*}
In Fig.~\ref{fig:Optimal_random_1-4_Der_1_MS3OP_mean_1} the results after applying dynamical ejections with the mean ejection fraction are plotted. 

In comparison to Fig.~\ref{fig:Optimal_random_1-4_ini}, the $m_\mathrm{max}$--$M_\mathrm{ecl}$ relation in the upper-left panel of Fig.~\ref{fig:Optimal_random_1-4_Der_1_MS3OP_mean_1} appears more dispersed but remains visible.
In particular, clusters older than $2\,\mathrm{Myr}$ with $M_\mathrm{cl} \approx 900$--$1700\,M_\odot$ show the largest changes relative to the initial plot.
This is consistent with the peak of the ejection fraction at this cluster-mass scale (see Fig.~2 in \citealt{Oh_Dependency_DynamicalEjections}). A higher ejection probability for the most massive star reduces $m_\mathrm{max}$  and, to a lesser extent, $M_\mathrm{cl}$, shifting clusters downwards and slightly leftwards. Note that the lower panels use different y-axis limits than in Fig.~\ref{fig:Optimal_random_1-4_ini}, which affects the visual impression.
For clusters with $M_\mathrm{cl}\gtrsim 10^{4.5}\,M_\odot$, we did not apply dynamical ejections because the fraction of ejected massive stars becomes negligible \citep{Oh_Dependency_DynamicalEjections}. We therefore leave the highest-mass model clusters unchanged with respect to dynamical ejections.

The  randomly sampled clusters change less, except that more objects populate the low-$M_\mathrm{cl}$ regime. These clusters are typically at $4\,\mathrm{Myr}$ and have undergone substantial mass loss and dynamical ejections, which reduces the remaining stellar mass reservoir. Optimally sampled clusters are less affected by stellar evolution and dynamical ejections because their stellar populations contain a larger fraction of low-mass stars, which are less likely to be ejected and lose less mass over these young ages.

The resulting values of the number of clusters, weighted mean and standard deviation are listed in Table~\ref{tab:Mean_Stdv_Der_mean}. 
Compared to before, the mass of the randomly sampled clusters is reduced because there are more clusters in the first bin than before and less in the second and third bins. Furthermore, one cluster is dissolved or has $M_\mathrm{cl}<10\,M_\odot$ because there are only 159 clusters in the fifth bin. The distribution of optimally sampled clusters does not change significantly.
Only one cluster moved from the third bin into the second. This supports the argumentation from before that the $M_\mathrm{cl}$ of the randomly sampled clusters is more sensitive to dynamical ejections because of the higher amount of massive stars.
The values of $\langle L_{\mathrm{H}\alpha} /M_\mathrm{cl} \rangle$ and $\sigma_{\langle L_{\mathrm{H}\alpha} /M_\mathrm{cl} \rangle}$ are still in the same range as before for the randomly and optimally sampled clusters. 
\begin{figure*}[t]
\sidecaption
        \includegraphics[width=12cm]{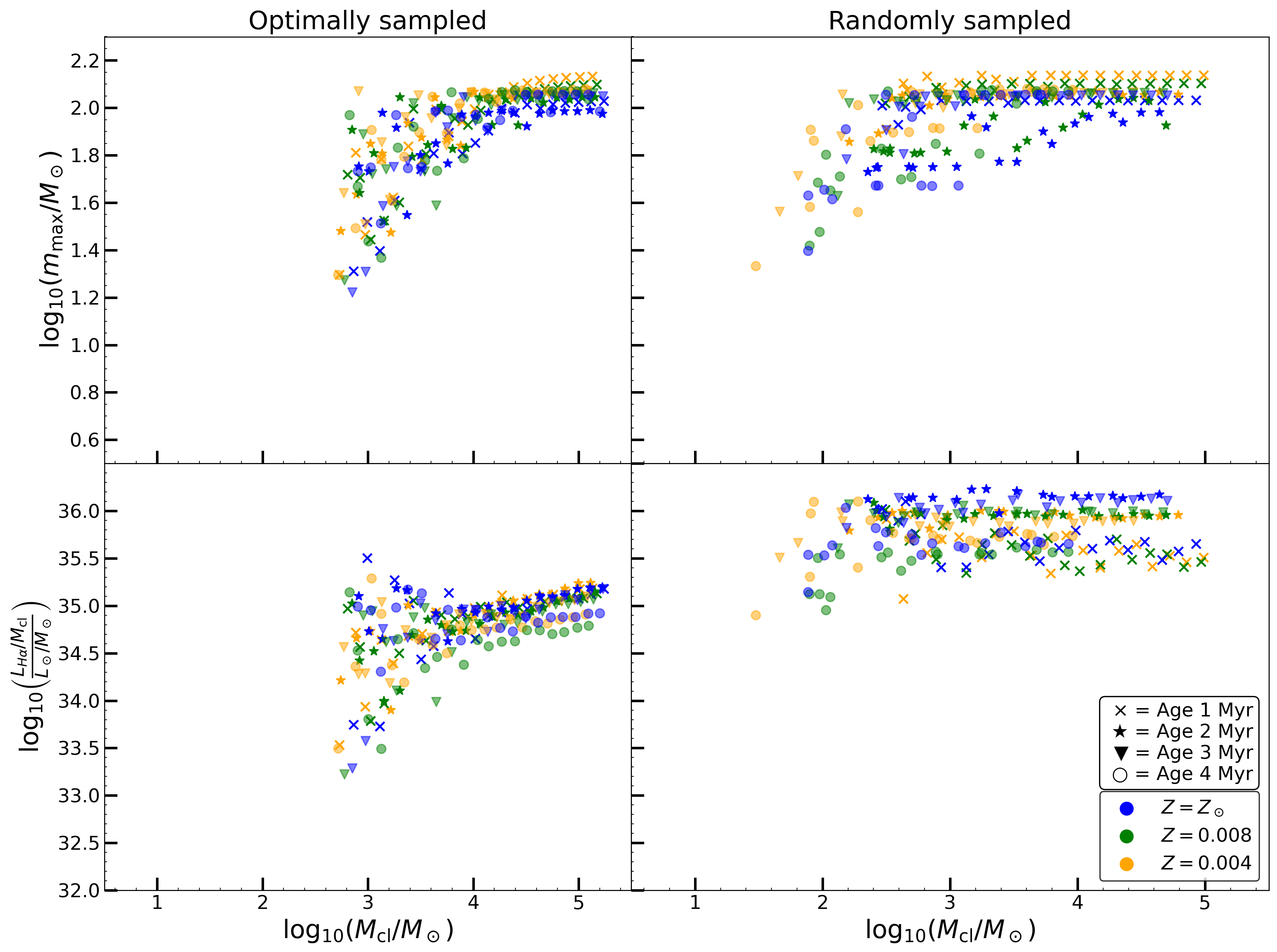} 
                \caption{Same as Fig.~\ref{fig:Optimal_random_1-4_ini} but including dynamical ejections with a mean ejection fraction, an age uncertainty of 1 Myr, and field stars.}
                \label{fig:Optimal_random_1-4_Der_1_MS3OP_mean_1_1_Fs}
\end{figure*}

\begin{figure*}[t]
\sidecaption
        \includegraphics[width=12cm]{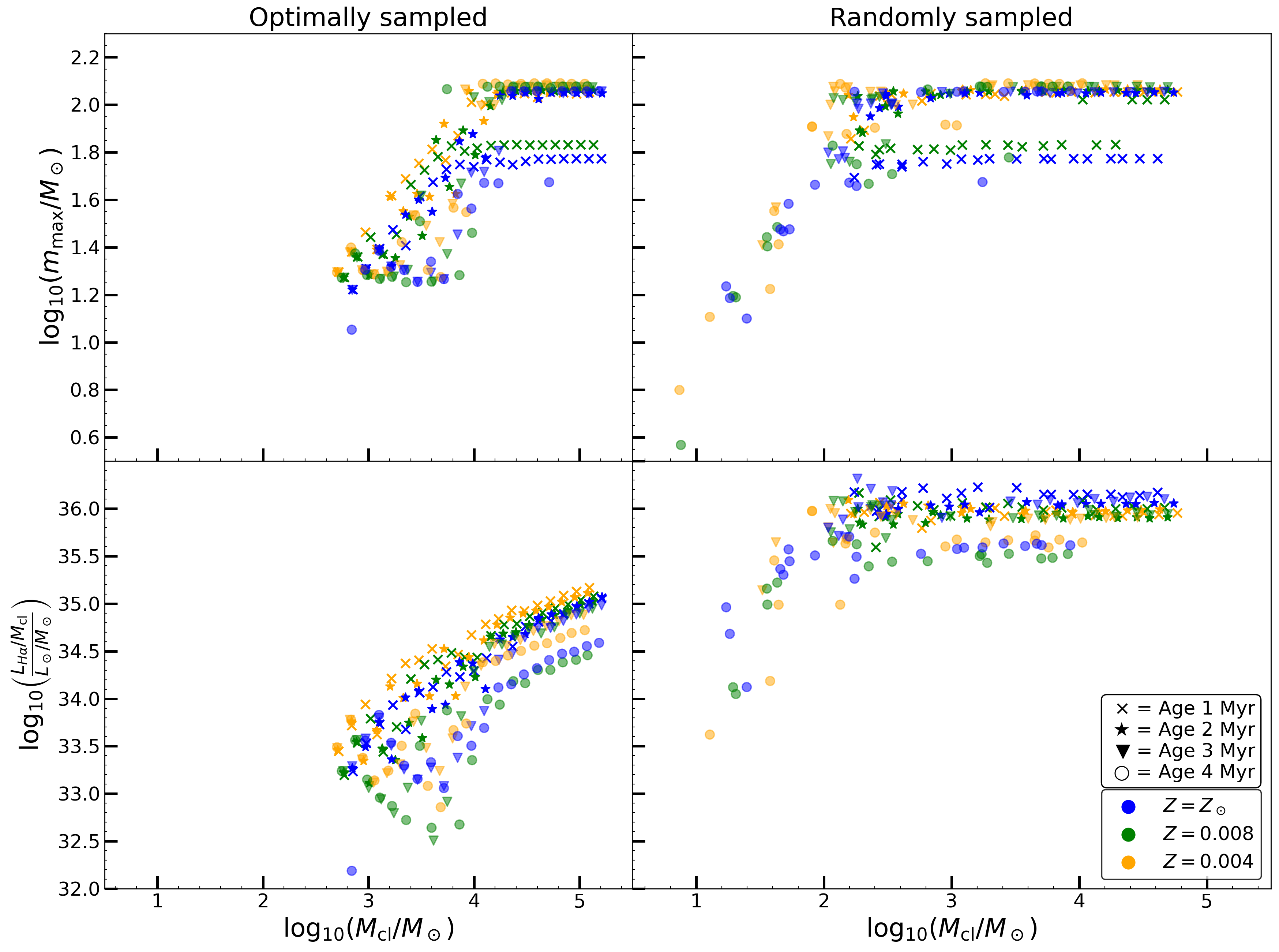} 
                \caption{Same as Fig.~\ref{fig:Optimal_random_1-4_ini} but including dynamical ejections with a $2\sigma$ higher ejection fraction and an age uncertainty of 2 Myr.}
                \label{fig:Optimal_random_1-4_Der_1_MS3OP_max_2_2}
\end{figure*}

\begin{figure*}[t]
\sidecaption
        \includegraphics[width=12cm]{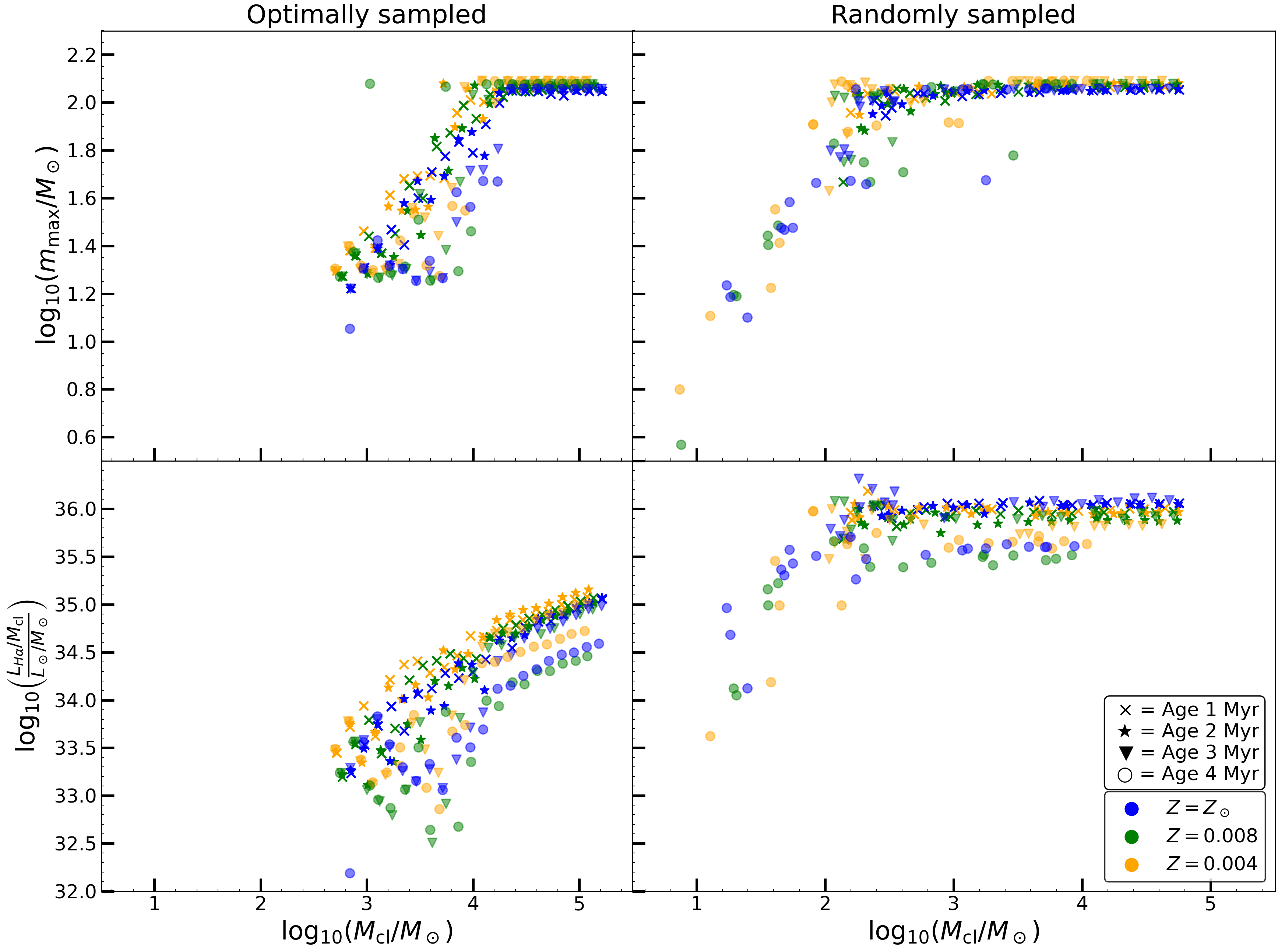} 
                \caption{Same as Fig.~\ref{fig:Optimal_random_1-4_ini} but including dynamical ejections with a $2\sigma$ higher ejection fraction and an age uncertainty of 3 Myr.}
                \label{fig:Optimal_random_1-4_Der_1_MS3OP_max_2_3}
\end{figure*}

\end{appendix}

\end{document}